\documentclass[english]{aa}
\usepackage[OT1]{fontenc}
\usepackage[latin9]{inputenc}
\usepackage{booktabs}
\usepackage{textcomp}
\usepackage{color}
\usepackage{graphicx}
\usepackage{amssymb}
\IfFileExists{url.sty}{\usepackage{url}}
                      {\newcommand{\url}{\texttt}}
\usepackage[authoryear]{natbib}

\makeatletter
\usepackage{rotating}
\usepackage{afterpage}
\usepackage[]{hyperref}
\usepackage{txfonts}
\hypersetup{
pdfauthor = {Konrad R. W. Tristram},
pdftitle = {Parsec-scale dust distributions in Seyfert galaxies}}
\usepackage[all]{hypcap}

\let\HyperRaiseLinkLength\@tempdima
\let\HyperRaiseLinkHook\@empty
\renewcommand{\HyperRaiseLinkDefault}{1.5\baselineskip}
\def\Hy@raisedlink#1{%
   \setlength\HyperRaiseLinkLength\HyperRaiseLinkDefault
   \HyperRaiseLinkHook
   \ifvmode
     #1%
   \else
     \penalty\@M
     \smash{\raise\HyperRaiseLinkLength\hbox{#1}}%
   \fi
 }

\providecommand*{\footref}[1]{%
  \begingroup
    \unrestored@protected@xdef\@thefnmark{\ref{#1}}%
  \endgroup
  \@footnotemark
}

\bibpunct{(}{)}{;}{a}{}{,}

\makeatother

\begin{document}

\title{Parsec-scale dust distributions in Seyfert galaxies\thanks{Based on Guaranteed Time Observations of the MIDI consortium collected at the European Southern Observatory, Chile, programme numbers 076.B-0038(A), 077.B-0026(B), 078.B-0031(A), 079.B-0180(A), 080.B-0258(A) and 081.D-0092(A).}}

\subtitle{Results of the MIDI AGN snapshot survey}

\author{K. R. W. Tristram\inst{1}, D. Raban\inst{2}, K. Meisenheimer\inst{3},
W. Jaffe\inst{2}, H. Röttgering\inst{2}, L. Burtscher\inst{3},
W. D. Cotton\inst{4}, U. Graser\inst{3}, Th. Henning\inst{3}, Ch.
Leinert\inst{3}, B. Lopez\inst{5}, S. Morel\inst{6}, G. Perrin\inst{7},
M. Wittkowski\inst{8}}

\institute{Max-Planck-Institut für Radioastronomie, Auf dem Hügel 69, 53121
Bonn, Germany \\
email: \url{tristram@mpifr-bonn.mpg.de}\and Leiden Observatory,
Leiden University, Niels-Bohr-Weg 2, 2300 CA Leiden, The Netherlands
\and Max-Planck-Institut für Astronomie, Königstuhl 17, 69117 Heidelberg,
Germany \and NRAO, 520 Edgemont Road, Charlottesville, VA 22903-2475,
USA \and Laboratoire H. Fizeau, UMR 6525, Université de Nice-Sofia
et Obs. de la Côte d'Azur, BP~4229, 06304 Nice Cedex~4, France.\and
European Southern Observatory, Casilla 19001, Santiago 19, Chile\and
LESIA, UMR 8109, Observatoire de Paris-Meudon, 5 place Jules Janssen,
92195 Meudon Cedex, France \and European Southern Observatory, Karl-Schwarzschild-Strasse
2, 85748 Garching bei München, Germany }

\titlerunning{Parsec-scale dust distribution in Seyfert galaxies}

\authorrunning{K. R. W. Tristram et al.}

\abstract{} {The emission of warm dust dominates the mid-infrared spectra
of active galactic nuclei (AGN). Only interferometric observations
provide the necessary angular resolution to resolve the nuclear dust
and to study its distribution and properties. The investigation of
dust in AGN cores is hence one of the main science goals for the MID-infrared
Interferometric instrument MIDI at the VLTI. As the first step, the
feasibility of AGN observations was verified and the most promising
sources for detailed studies were identified. } {This was carried
out in a {}``snapshot survey'' with MIDI using Guaranteed Time Observations.
In the survey, observations were attempted for 13 of the brightest
AGN in the mid-infrared which are visible from Paranal. } {The results
of the three brightest, best studied sources have been published in
separate papers. Here we present the interferometric observations
for the remaining 10, fainter AGN. For 8 of these, interferometric
measurements could be carried out. Size estimates or limits on the
spatial extent of the AGN-heated dust were derived from the interferometric
data of 7 AGN. These indicate that the dust distributions are compact,
with sizes on the order of a few parsec. The derived sizes roughly
scale with the square root of the luminosity in the mid-infrared,
$s\propto\sqrt{L_{\mathrm{MIR}}}$, with no clear distinction between
type 1 and type 2 objects. This is in agreement with a model of nearly
optically thick dust structures heated to $T\sim300\,\mathrm{K}$.
For three sources, the $10\,\mu\mathrm{m}$ feature due to silicates
is tentatively detected either in emission or in absorption. } {The
faint AGN of the snapshot survey are at the sensitivity limit of observations
with MIDI. Thus, the data set presented here provides a good insight
into the observational difficulties and their implications for the
observing strategy and data analysis. Based on the results for all
AGN studied with MIDI so far, we conclude that in the mid-infrared
the differences between individual galactic nuclei are greater than
the generic differences between type 1 and type 2 objects.}


\date{Received 30 December 2008 / Accepted 5 May 2009}

\maketitle

\section{Introduction\label{sec:introduction}}

In the standard model for active galactic nuclei (AGN), the central
engine, consisting of a hot accretion disk around a supermassive black
hole in the centre of a galaxy, is assumed to be surrounded by a geometrically
thick torus of obscuring gas and dust. These dusty tori are held responsible
for redistributing the optical and UV radiation from the hot accretion
disk into the mid- and far-infrared as well as for the type 1 / type
2 dichotomy of AGN. In type 2 AGN, the torus is thought to be oriented
edge-on so that it blocks the view towards the central engine; in
type 1 AGN the torus is oriented face-on, allowing a direct view towards
the central engine. The two types of AGN are hence considered to be
intrinsically the same with any differences in their appearance arising
from orientation effects \citep[for a review, see][]{1993Antonucci}.
However, this concept faces difficulties, as geometrically thick structures
orbiting compact objects are expected to collaps to a thin disk within
a few orbital periods \citep{1988Krolik}. Therefore, different theoretical
concepts for the nuclear dust distribution and obscuring medium are
currently being discussed. These include clumpy dusty tori \textendash{}
either supported by radiation pressure \citep{1992PierA,2007Krolik},
by elastic collisions of the clouds \citep{1988Krolik,2004Beckert}
or by supernova explosions \citep{2002Wada,2009Schartmann} \textendash{},
clumpy disk winds \citep[e.g.][]{1994Konigl,2006Elitzur} or warped
disks \citep[e.g.][]{2005Nayakshin}.

The dusty tori are too small to be significantly resolved with single
dish telescopes in the thermal infrared ($\lambda\gtrsim3\,\mu\mathrm{m}$),
even for $10\,\mathrm{m}$ class telescopes at the diffraction limit
(with or without AO), which implies that the dust distributions are
rather compact with sizes $\mathrm{\lesssim}10\,\mathrm{pc}$ \citep[e.g.][]{2003Soifer,2009Horst}.
Until recently, the principle arguments for the existence of these
tori were theoretical considerations and indirect observational evidence
such as the spectral energy distributions of AGN and the polarisation
properties of Seyfert 2 nuclei.%
\begin{table*}
\caption{Original list of targets for the MIDI SDT and GTO programme on AGN.
\label{table:target-list-original}}

\renewcommand{\footnoterule}{}

\begin{minipage}[c][1\totalheight][t]{1\textwidth}%
\centering

\begin{tabular}{llrrlrrccc}
\toprule 
\hline Galaxy & Alternative  &  &  &  & $D$ & $\Delta(1\,\mathrm{pc})$ & \multicolumn{2}{l}{Coudé guide star} & $F_{11.9\,\mathrm{\mu m}}$\tabularnewline
name & name & RA (J2000) & Dec (J2000) & type & {[}Mpc] & {[}mas] & $\psi$ {[}$''$] & $V$ {[}mag] & {[}Jy]\tabularnewline
(1) & (2) & (3) & (4) & (5) & (6) & (7) & \multicolumn{1}{c}{(8)} & (9) & (10)\tabularnewline
\midrule
\object{NGC~253} &  & $00\;47\;32.82$ & $-25\;17\;19.6$ & HII & 3 & 60\textcolor{white}{.0} & \multicolumn{2}{c}{none} & 2.04\tabularnewline
\object{NGC~1068} & M~77 & $02\;42\;40.70$ & $-00\;00\;48.0$ & Sy~2 & 14 & 14\textcolor{white}{.3} & 57 & 17 & \tabularnewline
\object{\textbf{NGC~1365}} &  & \textbf{$03\;33\;36.38$} & \textbf{$-36\;08\;25.7$} & \textbf{Sy~1.8} & \textbf{18} & \textbf{11}\textbf{\textcolor{white}{.3}} & \textbf{65} & \textbf{15} & \textbf{0.61}\tabularnewline
\object{\textbf{IRAS~05189-2524}} & \textbf{LEDA~17155} & \textbf{$05\;21\;01.40$} & \textbf{$-25\;21\;45.3$} & \textbf{Sy~2} & \textbf{170} & \textbf{1.2} & \multicolumn{2}{c}{\textbf{none}} & \textbf{0.55}\tabularnewline
\object{\textbf{MCG-05-23-016}} & \textbf{ESO~434-G040} & \textbf{$09\;47\;40.19$} & \textbf{$-30\;56\;56.4$} & \textbf{Sy~2} & \textbf{35} & \textbf{6}\textbf{\textcolor{white}{.0}} & \textbf{22} & \textbf{15} & \textbf{0.65}\tabularnewline
\object{\textbf{Mrk~1239}} &  & \textbf{$09\;52\;19.10$} & \textbf{$-01\;36\;43.5$} & \textbf{Sy~1.5} & \textbf{80} & \textbf{2.5} & \textbf{54} & \textbf{15} & \textbf{0.64}\tabularnewline
\object{NGC~3256} &  & $10\;27\;51.24$ & $-43\;54\;13.9$ & HII & 40 & 5\textcolor{white}{.3} & 32 & 15 & 0.55\tabularnewline
\object{\textbf{NGC~3281}} &  & \textbf{$10\;31\;52.10$} & \textbf{$-34\;51\;13.3$} & \textbf{Sy~2} & \textbf{45} & \textbf{5}\textbf{\textcolor{white}{.0}} & \textbf{22} & \textbf{16} & \textbf{0.63}\tabularnewline
\object{NGC~3783} &  & $11\;39\;01.71$ & $-37\;44\;19.0$ & Sy~1 & 40 & 5\textcolor{white}{.2} & 45 & 17 & 0.59\tabularnewline
\object{\textbf{NGC~4151}} &  & \textbf{$12\;10\;32.63$} & \textbf{$+39\;24\;20.7$} & \textbf{Sy~1.5} & \textbf{14} & \textbf{15}\textbf{\textcolor{white}{.1}} & \multicolumn{2}{c}{\textbf{none}} & \tabularnewline
\object{\textbf{3C~273}} &  & \textbf{$12\;29\;06.70$} & \textbf{$+02\;03\;08.6$} & \textbf{QSO} & \textbf{650} & \textbf{0.3} & \textbf{53} & \textbf{14} & \textbf{0.35}\tabularnewline
\object{Centaurus A} & NGC~5128 & $13\;25\;27.62$ & $-43\;01\;08.8$ & FR~I & 4 & 54\textcolor{white}{.7} & 44 & 13 & 1.22\tabularnewline
\object{\textbf{IC~4329A}} & \textbf{ESO~445-G050} & \textbf{$13\;49\;19.35$} & \textbf{$-30\;18\;34.4$} & \textbf{Sy~1.2} & \textbf{65} & \textbf{3.1} & \textbf{58} & \textbf{17} & \textbf{0.35}\tabularnewline
\object{Mrk~463E} &  & $13\;56\;02.89$ & $+18\;22\;18.2$ & Sy~1/2 & 200 & 1.0 & \multicolumn{2}{c}{none} & 0.34\tabularnewline
\object{Circinus} & ESO~097-G013 & $14\;13\;09.95$ & $-65\;20\;21.2$ & Sy~2 & 4 & 50\textcolor{white}{.0} & 50 & 12 & 9.70\tabularnewline
\object{\textbf{NGC~5506}} & \textbf{Mrk~1376} & \textbf{$14\;13\;14.88$} & \textbf{$-03\;12\;27.7$} & \textbf{Sy~1.9} & \textbf{25} & \textbf{8}\textbf{\textcolor{white}{.2}} & \multicolumn{2}{c}{\textbf{none}} & \textbf{0.91}\tabularnewline
\object{\textbf{NGC~7469}} & \textbf{Mrk~1514} & \textbf{$23\;03\;15.68$} & \textbf{$+08\;52\;26.3$} & \textbf{Sy~1.2} & \textbf{65} & \textbf{3.1} & \textbf{13} & \textbf{15} & \textbf{0.41}\tabularnewline
\object{NGC~7582} &  & $23\;18\;23.63$ & $-42\;22\;13.1$ & Sy~2 & 22 & 10\textcolor{white}{.0} & 54 & 17 & 0.67\tabularnewline
\bottomrule
\end{tabular}%
\end{minipage}%

~

\begin{minipage}[t][1\totalheight]{1\textwidth}%
Notes: The columns are: (1) the name of the galaxy used in this paper;
(2) an alternative name for the galaxy; (3) and (4) the coordinates
of the nucleus used for the MIDI observations or, if no observations
were carried out, the 2MASS coordinates; (5) the galaxy type (from
NED); (6) the distance to Earth assuming $H_{0}=73\,\mathrm{km}\,\mathrm{s}^{-1}\,\mathrm{Mpc}^{-1}$;
(7) the angular size corresponding to $1\,\mathrm{pc}$ in the target;
(8) the angular distance $\psi$ and (9) the visual magnitude $V$
of the best Coudé guide star (if present); (10) the flux at $11.9\,\mathrm{\mu m}$
from \citealt{2008Raban}.%
\end{minipage}%

\end{table*}

The instrumental means to spatially resolve the nuclei of active galaxies
at infrared wavelengths became available only with the advent of sufficiently
sensitive interferometric instruments. One of these is the MID-infrared
Interferometric instrument (MIDI) at the Very Large Telescope Interferometer
(VLTI), located on Cerro Paranal in northern Chile and operated by
the European Southern Observatory (ESO). Apart from two observations
in the near-infrared with other interferomers, namely of the Seyfert
1 galaxy NGC~4151 with the Keck interferometer \citep{2003Swain}
and of the Seyfert 2 galaxy NGC~1068 with the VINCI instrument also
at the VLTI \citep{2004Wittkowski}, MIDI is currently the only interferometer
to successfully observe active galactic nuclei in the infrared wavelength
range.

The first AGN to be targeted by MIDI was the brightest AGN in the
mid-infrared, NGC~1068. These observations were carried out during
the Science Demonstration Time (SDT) of the instrument and succeded
in resolved the nuclear dust emission \citep{2004Jaffe1,2006Poncelet}.
This demonstrated the great potential of MIDI for the investigation
of AGN. Further MIDI observations of NGC~1068 were subsequently carried
out in Open Time \citep{2009Raban}.

Using mainly Guaranteed Time Observations (GTO), the two next brightest
AGN, the radio galaxy Centaurus~A and the Circinus galaxy, a Seyfert
2 galaxy, were studied with MIDI. The results were published in \citet{2007Meisenheimer}
and \citet{2007Tristram2} respectively. The MIDI observations of
NGC~1068 and the Circinus galaxy have directly confirmed that, indeed,
compact AGN heated dust structures on the scale of a few parsecs exist
in Seyfert 2 galaxies and that it is possible to determine their size
as well as their orientations with respect to the source axis (\citealp{2004Jaffe1,2006Poncelet,2007Tristram2}
and \citealp{2009Raban}). On the other hand, the mid-infrared flux
from Centaurus A is dominated by an unresolved core (most likely the
high-frequency tail of the radio core) which is surrounded by a low
luminosity {}``dust disk'' of about $0.6\,\mathrm{pc}$ diameter
\citep{2007Meisenheimer}.

Apart from these three well studied sources, all other AGN which can
potentially be observed with MIDI are less bright in the mid-infrared
and are close to, or at, the sensitivity limit of MIDI. In order to
determine which AGN are suitable for further study with MIDI, a snapshot
survey of potential targets was carried out during GTO. In this survey,
each source on the target list was to be observed at least once in
order to test for the feasibility of MIDI observations as well as
to obtain first estimates for the basic mid-infrared properties of
the target sources. The results of this survey are presented in this
paper.

The paper is organised as follows: Sect.~\ref{sec:target-list} describes
the original target list together with the properties of the sources;
in Sect.~\ref{sec:observations}, the general observational strategy
as well as the general challenges for the data reduction are presented;
in Sect.~\ref{sec:results}, the results for the individual sources
are discussed, while in Sect.~\ref{sec:summary_observations} a summary
of the observational results is given. The discussion and the conclusions
are found in Sects.~\ref{sec:discussion} and \ref{sec:conclusions}
respectively. A more detailed discussion of the observations and the
data reduction procedures for the individual sources is given in Appendix~\ref{sec:observations_individual}.

\section{Target list\label{sec:target-list}}

The original list of targets for both the GTO and the SDT (see Table~\ref{table:target-list-original})
contains those 16 active galaxies from \citet{2008Raban}, plus NGC~1068,
which are well observable from Cerro Paranal ($\mathrm{Dec}<+25^{\circ}$)
and which were thought to be bright enough for MIDI observations (unresolved
core flux $F_{\mathrm{core}}\gtrsim0.35\,\mathrm{Jy}$). Additionally,
the prototypical Seyfert 1 galaxy NGC~4151 was included in the list.
Because of its northerly position at $\mathrm{Dec}=+39^{\circ}\,24'$,
the latter does not rise more than $26^{\circ}$ (corresponding to
an airmass of 2.3) above the horizon at the location of the VLTI ($24^{\circ}\,40'\,\mathrm{S}$).
Despite the difficulty of observing it from Paranal, it was also included
in the list of GTO targets due to (1) its importance as the nearest
and brightest type 1 AGN and of (2) it being one of the two AGN observed
by other interferometers.%
\begin{table*}
\caption{Overview of the observations and observation attempts for the sources
from Table \ref{table:target-list-original}.\label{table:target-list-observations}}

\renewcommand{\footnoterule}{}

\begin{minipage}[c][1\totalheight][t]{1\textwidth}%
\centering

\begin{tabular}{r@{ }lllcrcl}
\toprule 
\hline  & Galaxy & First / last & Programme & Coudé & \multicolumn{2}{c}{Fringe} & Result of the MIDI observations\tabularnewline
 & name & observation~~~~~~~~~ & number & guiding & \multicolumn{2}{c}{tracks} & \tabularnewline
 & (1) & (2) & (3) & (4) & \multicolumn{2}{c}{(5)} & ~~~~~~~~~~~~~~~~~(6)\tabularnewline
\midrule
 & NGC~253 & \multicolumn{3}{l}{observed by S. Hönig (no fringes detected)} & ~~~~~~~~~~ &  & \tabularnewline
 & NGC~1068 & 2003 Jun 15 & 060.A-9224(A)%
\footnote{\label{fn:footlabel_a}The first observations of NGC~1068 and the
Circinus galaxy were carried out in SDT.%
} & nucleus & 9%
\footnote{\label{fn:footlabel_b}In the meantime, more observations than those
listed here have been obtained for several of the original SDT and
GTO targets in Open Time.%
} &  & well resolved \citep{2009Raban}\tabularnewline
\textbf{{*}} & \textbf{NGC~1365} & \textbf{2006 Sep 11} & \textbf{077.B-0026(B)} & \textbf{nucleus} & \textbf{7\hspace{1.23mm}} &  & \textbf{partially resolved (possibly elongated)}\tabularnewline
\textbf{{*}} & \textbf{IRAS~05189-2524} & \textbf{2006 Sep 11} & \textbf{077.B-0026(B)} & \textbf{nucleus} & \textbf{1\hspace{1.23mm}} &  & \textbf{faint fringe detection only}\tabularnewline
\textbf{{*}} & \textbf{MCG-05-23-016} & \textbf{2005 Dec 19} & \textbf{076.B-0038(A)} & \textbf{star} & \textbf{1\footref{fn:footlabel_b}} &  & \textbf{partially resolved}\tabularnewline
\textbf{{*}} & \textbf{Mrk~1239} & \textbf{2005 Dec 19} & \textbf{076.B-0038(A)} & \textbf{nucleus} & \textbf{2\footref{fn:footlabel_b}} &  & \textbf{essentially unresolved}\tabularnewline
 & NGC~3256 & \multicolumn{4}{l}{} &  & not yet observed\tabularnewline
\textbf{{*}} & \textbf{NGC~3281} & \textbf{2007 Feb 07} & \textbf{078.B-0031(A)} & \textbf{nucleus} & \multicolumn{2}{c}{\textbf{none}} & \textbf{no AO correction possible using the nucleus}\tabularnewline
 & NGC~3783 & \multicolumn{4}{l}{see \citet{2008Beckert}} &  & partially resolved\tabularnewline
\textbf{{*}} & \textbf{NGC~4151} & \textbf{2007 Feb 07} & \textbf{078.B-0031(A)} & \textbf{nucleus} & \textbf{2\hspace{1.23mm}} &  & \textbf{well resolved}\tabularnewline
\textbf{{*}} & \textbf{3C~273} & \textbf{2007 Feb 07} & \textbf{078.B-0031(A)} & \textbf{nucleus} & \textbf{4\hspace{1.23mm}} &  & \textbf{possibly resolved}\tabularnewline
 & Centaurus~A & 2005 Mar 01 & 074.B-0213(B) & star & 5\footref{fn:footlabel_b} &  & partially resolved \citep{2007Meisenheimer}\tabularnewline
\textbf{{*}} & \textbf{IC~4329A} & \textbf{2007 Feb 07} & \textbf{078.B-0031(A)} & \textbf{nucleus} & \textbf{2\footref{fn:footlabel_b}} &  & \textbf{unresolved}\tabularnewline
 & Mrk~463E & \multicolumn{4}{l}{} &  & not yet observed\tabularnewline
 & Circinus & 2004 Jun 03 & 060.A-9224(A)\footref{fn:footlabel_a} & star & 22\footref{fn:footlabel_b} &  & well resolved \citep{2007Tristram2}\tabularnewline
\textbf{{*}} & \textbf{NGC~5506} & \textbf{2007 Feb 07} & \textbf{078.B-0031(A)} & \textbf{nucleus} & \multicolumn{2}{c}{\textbf{none}} & \textbf{no AO correction possible using the nucleus}\tabularnewline
\textbf{{*}} & \textbf{NGC~7469} & \textbf{2006 Sep 11} & \textbf{077.B-0026(B)} & \textbf{nucleus} & \textbf{3\hspace{1.23mm}} &  & \textbf{well resolved}\tabularnewline
 & NGC~7582 & \multicolumn{4}{l}{observed by S. Hönig (no fringes detected)} &  & \tabularnewline
\bottomrule
\end{tabular}%
\end{minipage}%

~

\begin{minipage}[t][1\totalheight]{1\textwidth}%
Notes: The objects of the snapshot survey discussed in this paper
are highlighted by boldface and marked by asterisks. The columns are:
(1) the name of the galaxy (repeated from Table~\ref{table:target-list-original});
(2) the date and (3) ESO programme number of the first successful
observation or the last unsuccessful observation attempt; (4) the
type of Coudé guiding used for the observations; (5) the number of
successful fringe tracks obtained in SDT and GTO (without judgement
on whether the data are useful or not) and (6) the result of the MIDI
observations.%
\end{minipage}%

\end{table*}

In preparation for the MIDI observations, high resolution imaging
was carried out for most of the AGN on the target list in the near-infrared
with NACO, the adaptive optics assisted multi mode instrument of the
Very Large Telescope \citep{2003Rousset,2003Lenzen}, and in the mid-infrared
with TIMMI2, the Thermal Infrared Multi-Mode Instrument 2 \citep{2003Kaeufl}.
Essentially, all nuclei remain unresolved both in the near- and mid-infrared.
The results of the investigation with TIMMI2 were published in \citet{2008Raban};
those of the investigation with NACO will be presented in \citet{2009Prieto}.
The photometric values of the targets at $11.9\,\mathrm{\mu m}$ from
\citet{2008Raban}, for aperture sizes ranging from $1.2\,\mathrm{arcsec}$
to $2.6\,\mathrm{arcsec}$, are listed in the last column of Table~\ref{table:target-list-original}.

As the observations progressed, several galaxies from the original
target list were released from the GTO protection without being observed.
NGC~253, NGC~3256, NGC~3783, Mrk~463E, and NGC~7582 are among
these galaxies released. For some of these sources, observations were
attempted by other authors: NGC~3783 was successfully observed in
Open Time by \citet{2008Beckert}; for NGC~253 and NGC~7582 no interferometric
signal could be detected (Hönig, private comm.).

A summary of the observations and results for all the sources from
the original target list is given in Table~\ref{table:target-list-observations}.
The ten objects discussed in detail in this paper, that is, the sources
of the snapshot survey, are highlighted by boldface and marked by
an asterisk.

\section{Observations and data reduction\label{sec:observations}}

\subsection{The MIDI instrument}

MIDI is a two beam Michelson type interferometer producing dispersed
fringes in the N band over a wavelength range from $8$ to $13\,\mu\mathrm{m}$
\citep{2003Leinert2}. For our observations, the light of two of the
$8.2\,\mathrm{m}$ unit telescopes (UTs) was combined at any one time.
The incoming wavefronts from the two telescopes were corrected using
the Multi Application Curvature Adaptive Optics system (MACAO, \citealp{2003Arsenault}).
All of the AGN in the snapshot survey were observed in high sensitivity
(HIGH-SENS) mode and with low spectral resolution ($\lambda/\delta\lambda\approx30$).
This mode uses a NaCl prism after the beam combiner. As the main goal
of this survey was to test for the feasibility of MIDI observations,
the two shortest baselines, UT2~--~UT3 and UT3~--~UT4, were used
for most of the observing runs to obtain the highest possible correlated
fluxes. An observation run planned for 2007 May 05 with the longer
UT2~--~UT4 baseline was completely lost due to bad weather conditions.

\subsection{Observing sequence}

We now briefly discuss the observing sequence and data reduction,
while putting a special emphasis on the challenges in the context
of sources at the detection limit of MIDI and the VLTI. A detailed
description of the observing sequence can be found in \citet{2007Tristram1}.
The peculiarities of the observations and data reduction for the individual
objects are discussed in Appendix \ref{sec:observations_individual}.

After the telescopes have been pointed at the source, the AO system,
MACAO, must close the loop on its reference source. This can be either
the nucleus of the source itself, or -- if available -- on a Coudé
guide star with $V<17\,\mathrm{mag}$ and within $57\,\mathrm{arcsec}$
of the source. For most of the sources this correction used the nucleus
itself (see Table \ref{table:target-list-observations}). In several
cases, this turned out to be at the limit of the abilities of the
MACAO system resulting in difficulties obtaining a stable correction.
This was especially true for poor seeing conditions.

The difficulty is largely that the AO corrections are determined in
the optical where the nuclei may be faint due to obscuration; in addition,
the optical cores of the host galaxies are spatially extended. This
issue will be further discussed in Sect.~\ref{sec:summary_observations}.

\begin{figure}
\centering

\includegraphics{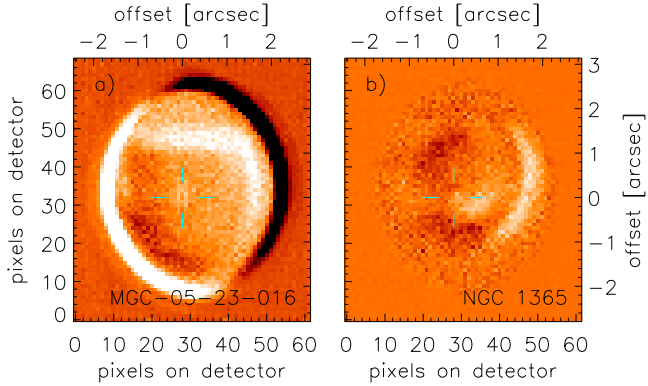}

\caption{Final acquisition images of \textbf{a)} MCG-05-23-016 (observed 2005
Dec 19) and \textbf{b)} NGC~1365 (observed 2006 Sep 11). Both sources
have a total flux of about $F\sim0.5\,\mathrm{Jy}$ and $\mathrm{NDIT}=16\,000$
was used. The sources are barely recognisable due to the high number
of frames. The residual background from the VLTI tunnels is much stronger
than the sources. A less even background would completely prevent
a detection of the source.}

\label{fig:observ_acquisition}
\end{figure}

The next step in the observing sequence is to align the beams from
the two telescopes. This uses acquisition images made with each telescope.
For our observations, the number of individual exposures had to be
increased significantly from the standard value of $\mathrm{NDIT}=1000$
to, for instance, $\mathrm{NDIT}=16\,000$ in order to make the source
visible against the residual background. Even for such increased integration
times, several of our sources were barely discernible against the
residual background, as can be seen from Fig.~\ref{fig:observ_acquisition}.%

Since our observations in September 2006, the VLTI infrared field-stabiliser
IRIS (InfraRed Image Sensor, \citealp{2004Gitton}) has been available.
The purpose of IRIS is to stabilise the field of the beams by measuring
the low-frequency tip-tilt directly in the VLTI laboratory, where
MIDI is located. For sources bright enough in the near-infrared ($K<14\,\mathrm{mag}$),
IRIS guarantees the correct alignment of the beams during the observations.
If one trusts that a proper alignment of the beams in IRIS also produces
a proper alignment of the beams in MIDI%
\footnote{The position of the photocentre of the source may, however, depend
on the wavelength. Also, at high airmasses, the wavelength dependency
of the atmospheric refraction leads to different apparent positions
of the source in the near and mid-infrared which might lead to slightly
different positions in MIDI and IRIS.%
}, no acquisition images with MIDI itself are necessary. Nevertheless,
we continued to record acquisition images in later observing runs
in order to verify the correct overlap of the beams and the quality
of the source image with respect to the background emission. Only
for two of the sources observed on 2007 Feb 07 (3C~273 and IC~4329A)
and for the last observation of NGC~1365 on 2007 Nov 24, no acquisition
images were obtained.

After the beams are aligned, the beamsplitter, a $0.52\,\mathrm{arcsec}$
wide slit and the prism are inserted into the light path and a so-called
\emph{fringe search} is performed in order to determine the position
of the zero optical path difference (OPD) between the two beams. If
an interferometric signal was detected in the fringe search, the path
difference is stabilised (except for a small modulation) at or near
zero OPD, while the spectrally dispersed interferometric signal is
continuously recorded in order to increase the signal to noise ratio.
This is the so-called \emph{fringe track}. Because the strong background
emission in the mid-infrared is uncorrelated, it can be effectively
removed during the data reduction process by high pass filtering of
the modulated fringe signal in delay and frequency space. Therefore,
no chopping needs to be performed during the interferometric measurements
and a more stable AO correction is achieved. All of our observations
were carried out in {}``offset tracking'' mode with the OPD stabilised
at $50\,\mu\mathrm{m}$ from zero OPD. This tracking mode is advantageous
for sky removal using a high pass filter in frequency space, as the
signal is always modulated in wavelength; at zero OPD this is not
the case.

Finally, photometric data, that is, single dish spectra, are recorded
using only one telescope at a time for an otherwise identical optical
set-up. As for the aquisition, chopping is used to suppress the background.

The entire procedure of acquisition, fringe search, fringe track and
photometry is repeated for the calibrator star to obtain a single
calibrated visibility point.

\subsection{Data reduction}

The data reduction was performed with EWS (Expert Work Station%
\footnote{The software package {}``MIA+EWS'' is available for download at
\url{http://www.strw.leidenuniv.nl/~nevec/MIDI/index.html}%
}, Version 1.7), a data reduction package based on the coherent data
reduction method. A detailed description of EWS is found in \citet{2004Jaffe2}.
The parameters for the EWS data reduction were $\mathtt{smooth}=10$,
$\mathtt{gsmooth}=20$, $\mathtt{dave}=1$ and $\mathtt{nophot}=1$,
the last parameter meaning that we chose to directly calibrate the
correlated fluxes without using the photometric fluxes. The calibration
used the database of stellar spectra from Roy van Boekel (private
comm.).

During the data reduction process, we noticed a secular drift of the
position of the dispersed signal on the MIDI detector. For data obtained
in November 2007, the shift had become so large ($\mathrm{\sim}3\,\mathrm{pixels}=0.26\,\mathrm{arcsec}$)
that the dispersed fringes and the spectra of the photometry were
moving outside the standard EWS mask. For this reason, we constructed
a separate mask for every observing run, taking into account the position
and width in spatial direction of the spectra. Additionally, the sky
bands used for the determination of the photometry were adjusted according
to the new mask position. 

The results of the data reduction process are the spectrally dispersed
correlated flux $F_{\mathrm{cor}}(\lambda)$ (from the interferometric
measurement), the total flux spectrum $F_{\mathrm{tot}}(\lambda)$
(from the photometric measurement, corresponding to an aperture of
$0.52\times1.1\,\mathrm{arcsec}^{2}$) and the visibility spectrum
$V(\lambda)$ (derived from both the interferometric and the photometric
data). For an ideal measurement, $V=F_{\mathrm{cor}}/F_{\mathrm{tot}}$.
In practice, however, this is often not the case (for a detailed discussion,
we refer to \citealt{2007Tristram1}). 

For sources as weak as those discussed here, the statistical errors
determined by EWS have to be considered with caution. Especially,
the photometry can have unrealistically large errors. These come from
the variations of the flux induced by the background residual. Extremely
large errors were simply truncated corresponding to a relative error
of $90\,\%$. The large errors of the photometry propagate into the
visibility, but not into the correlated flux due to its direct calibration.
The latter seems to have rather too small errors.

Due to the low quality or the incompleteness of several data sets,
especially the photometric components, special data reduction procedures
had to be applied to obtain meaningful results. The photometric data
were completed or their quality improved by mainly using the three
following methods: (1) where no photometry observations were carried
out at all, photometry data obtained in conjunction with another fringe
track were used to complete the data set; (2) if only one of the two
photometric measurements contains a useful signal (e.g. if there was
no AO correction for one of the two telescopes during chopping), only
that photometry was used to determine the total flux and (3) for strong
background residuals, only those portions of the photometry integration
where the background residual is relatively flat were used for the
determination of the total flux spectrum. A description of which method
and which additional steps were carried out for each of the individual
sources is given in Appendix~\ref{sec:observations_individual}.

\subsection{Compilation of comparison spectra}

In order to check whether the MIDI data are reasonable and in order
to be able to interpret the correlated fluxes where no or only a questionable
total flux spectrum was observed, additional mid-infrared spectra
were compiled. Preference was given to data obtained with the highest
angular resolution. For lack of alternative (no high resolution ground-based
spectra available), Spitzer spectra were retrieved from the archive
for several sources. The Spitzer spectra used for our comparison were
observed with the Short-Low (SL, $\lambda/\delta\lambda\approx60-120$)
module of IRS, the InfraRed Spectrograph \citep{2004Houck} on the
Spitzer Space Telescope, in staring mode. The nuclear spectra directly
delivered by the pipeline, version S15.3, were used, without applying
further correction factors. Note that a direct comparison of the Spitzer
IRS spectra with MIDI data is only possible with reservations due
to the large difference in the apertures of the two instruments ($3.6\times6.0\,\mathrm{arcsec}^{2}$
for IRS and $0.5\times1.1\,\mathrm{arcsec}^{2}$ for the total flux
in MIDI). This is especially the case for sources also harbouring
a nuclear starburst and thus displaying extended emission in the infrared,
such as NGC~7469 (see Sect.~\ref{sub:result_ngc7469}). The Spitzer
spectra thus have to be seen as upper limits to the ones measured
by MIDI.

\section{Results\label{sec:results}}

The results of the measurements are presented for those seven sources
where useful data could be obtained. These are NGC~1365, MCG-05-23-016,
Mrk~1239, NGC~4151, 3C~273, IC~4329A and NGC~7469. For the remaining
three sources of the snapshot survey, no useful data could be obtained:
for NGC~3281 and NGC~5506, no measurements with MIDI were possible
at all, because no stable AO correction could be achieved with MACAO
using the extended nuclei for guiding. For IRAS~05189-2524 the interferometric
signal was too weak for fringe tracking (see Appendix~\ref{sub:observ_leda17155}).

For those sources where both an interferometric measurement as well
as a proper measurement of the total flux spectrum was carried out
with MIDI, a very crude estimate or a limit on the size of the emission
region can be derived assuming a Gaussian brightness distribution.
A Gaussian flux distribution is the simplest and most general initial
guess for the shape of the source. Even if the true shape of the flux
distribution differs from a Gaussian, it provides us with a correct
scale of the spatial extent of the source.

The full width at half maximum $\mathit{FWHM}(\lambda)$ of the Gaussian
brightness distribution is related to the visibility $V(\lambda)$
and the projected baseline length $\mathit{BL}$ by\begin{equation}
\mathit{FWHM}(\lambda)=\frac{\lambda}{\mathit{BL}}\cdot\frac{2}{\pi}\sqrt{-\ln2\cdot\ln V(\lambda)}.\label{eq:faint_gauss-fwhm}\end{equation}
 By definition, $\mathit{FWHM}(\lambda)=0$ when $V(\lambda)=1$%
\footnote{Theoretically, $V(\lambda)$ cannot be greater than 1. Practically,
however, it is not uncommon to obtain $V(\lambda)\geq1$ simply due
to the error of the measurements. For such cases we set $\mathit{FWHM}=0$.%
}. Because of the nonlinear dependency of the $\mathit{FWHM}$ on the
visibility in Eq.~\ref{eq:faint_gauss-fwhm}, the upper and lower
limits on the $\mathit{FWHM}$ were calculated directly from the upper
and lower limits of the visibility.

\subsection{NGC~1365\label{sub:result_ngc1365}}

The large, barred spiral galaxy NGC~1365 is located at a distance
of $18.3\pm3.3\,\mathrm{Mpc}$ ($1\,\mathrm{arcsec}=90\,\mathrm{pc}$,
\citealt{1999Silbermann}) in the Fornax cluster of galaxies. The
Seyfert type of this galaxy is variable. This is seen as prime evidence
for a clumpy torus \citep[e.g.][and references therein]{2009Risaliti}.
An extensive overview of this galaxy is found in \citet{1999Lindblad}.
In the mid-infrared, TIMMI2 data show that the nucleus is surrounded
by 7 individual sources within $20\,\mathrm{arcsec}$ from the nucleus
\citep{2005Galliano,2008Raban}, which are interpreted as embedded
young star clusters. The nuclear source itself shows a slight extension
in E-W direction in the data from \citeauthor{2005Galliano}, but
this extension has not been confirmed by \citet{2008Raban}. A TIMMI2
spectrum of the nucleus is presented in \citet{2004Siebenmorgen}.
It is featureless and flat, without any evidence of silicate absorption
or emission.

\begin{figure*}[t]
\centering

\includegraphics{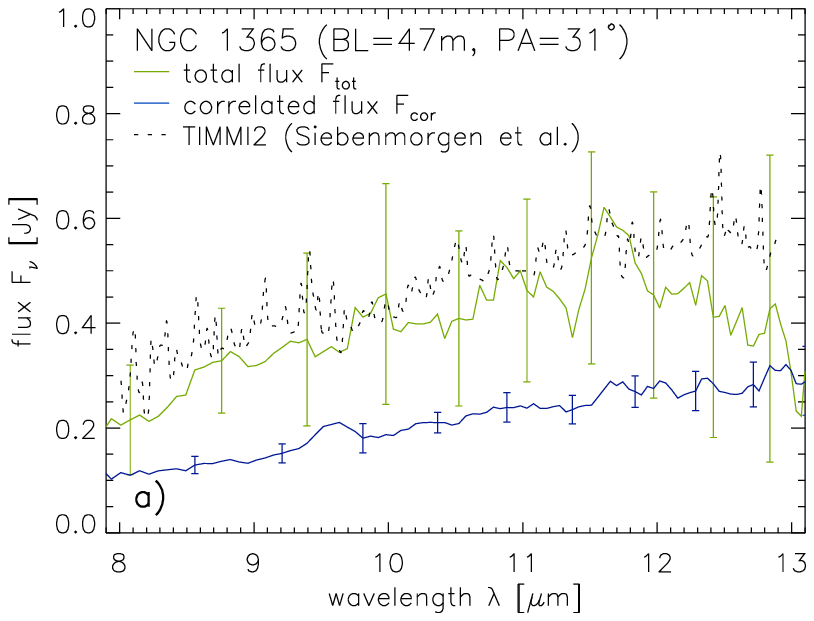}\hfill\includegraphics{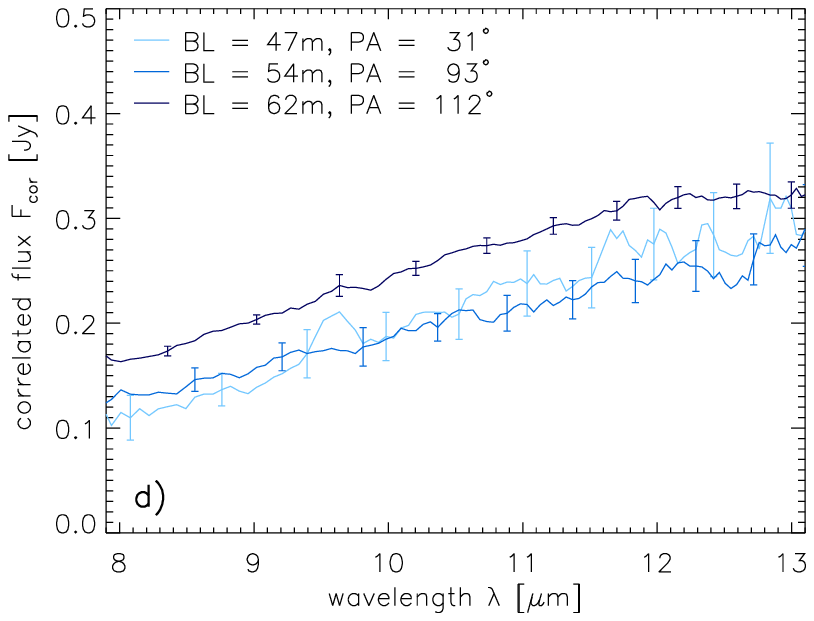}

\includegraphics{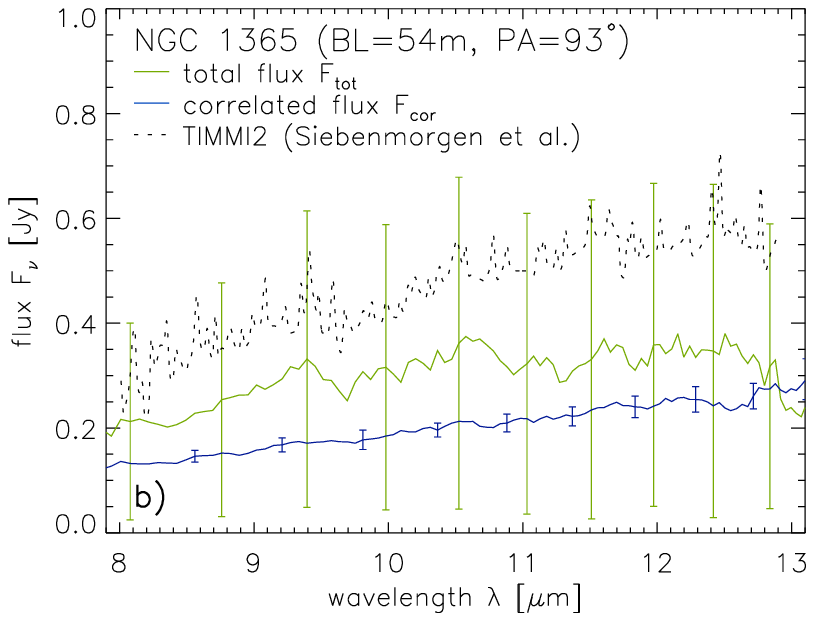}\hfill\includegraphics{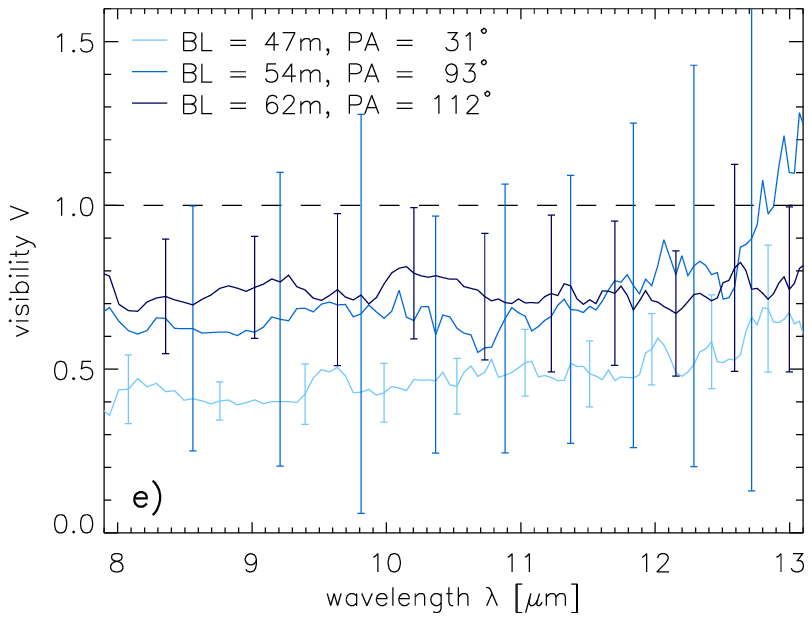}

\includegraphics{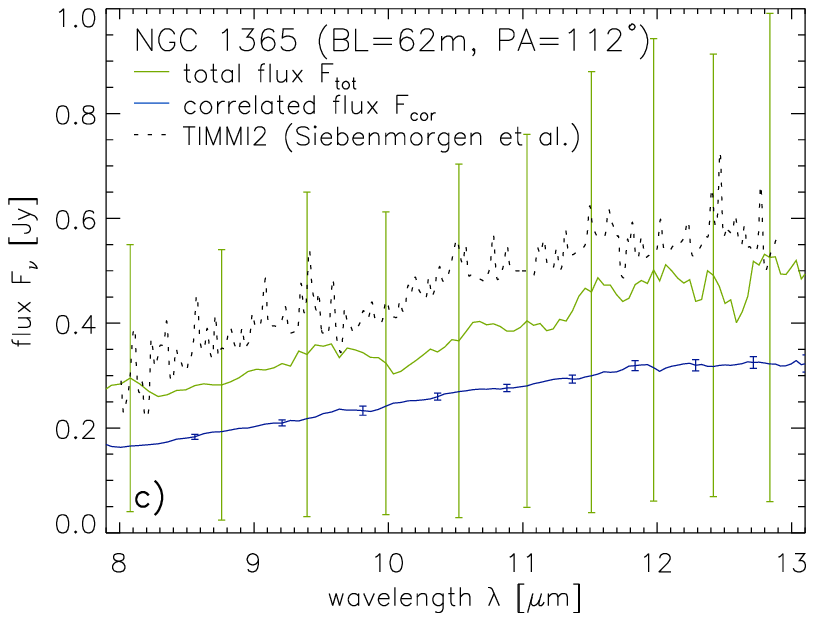}\hfill\includegraphics{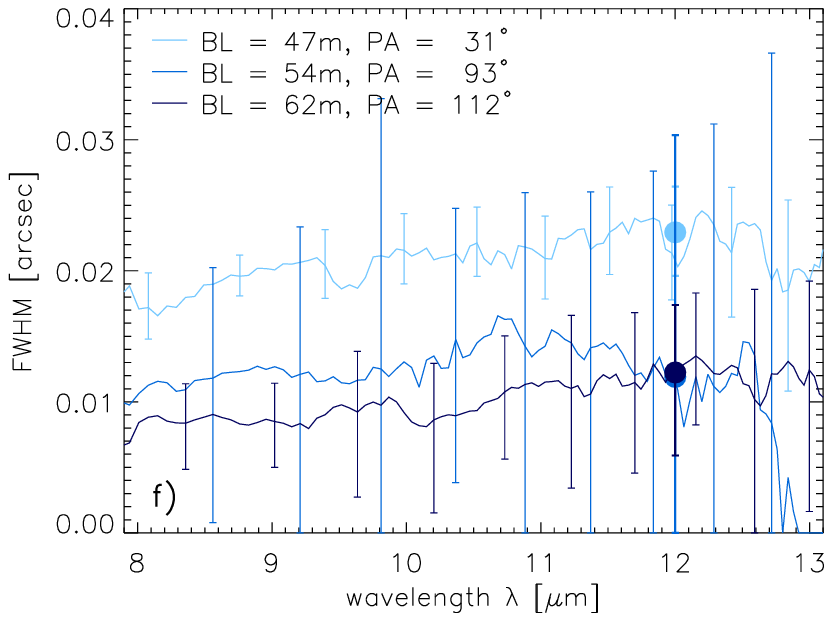}

\caption{Correlated and total flux spectra, visibilities and size estimates
for the three interferometric measurements of NGC~1365. For sake
of clarity, the error bars for the MIDI data are given for every tenth
wavelength bin. In \textit{the left column}, panels \textbf{a)}--\textbf{c)},
the correlated (blue) and total (green) flux spectra are displayed
for each of the three baseline orientations. For comparison, the TIMMI2
spectrum from \citet{2004Siebenmorgen} is also shown (dotted line).
In panel \textbf{d)} the three measurements of the correlated flux
are shown together in a single plot. Panels \textbf{e)} and \textbf{f)}
show the visibility spectra and the derived $\mathit{FWHM}$ of a
corresponding Gaussian flux distribution, respectively. The size estimates
at $12\,\mathrm{\mu m}$ used in Sect.~\ref{sub:discussion_sizes}
are indicated in panel \textbf{f)} by filled circles.}
\label{fig:result_ngc1365}
\end{figure*}
\afterpage{\clearpage}

\begin{figure*}
\centering

\includegraphics{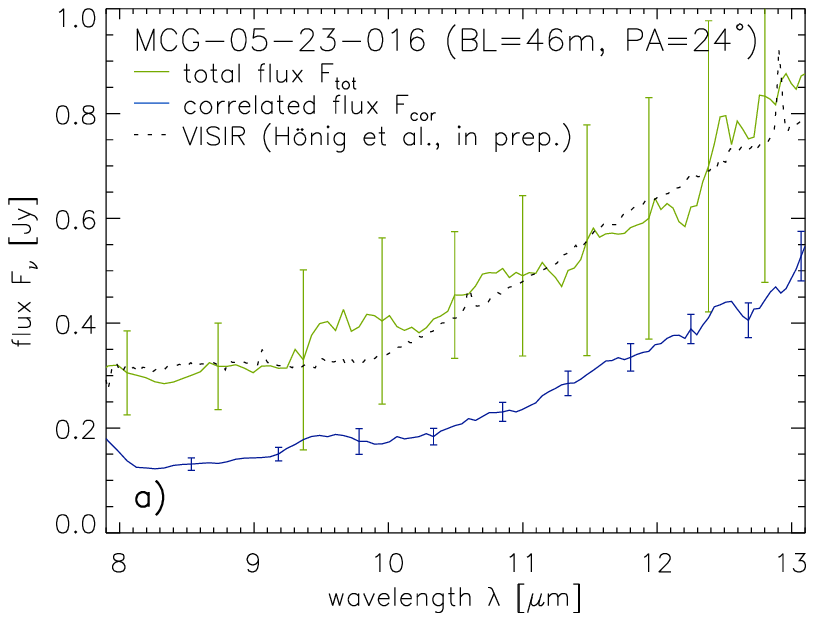}\hfill\includegraphics{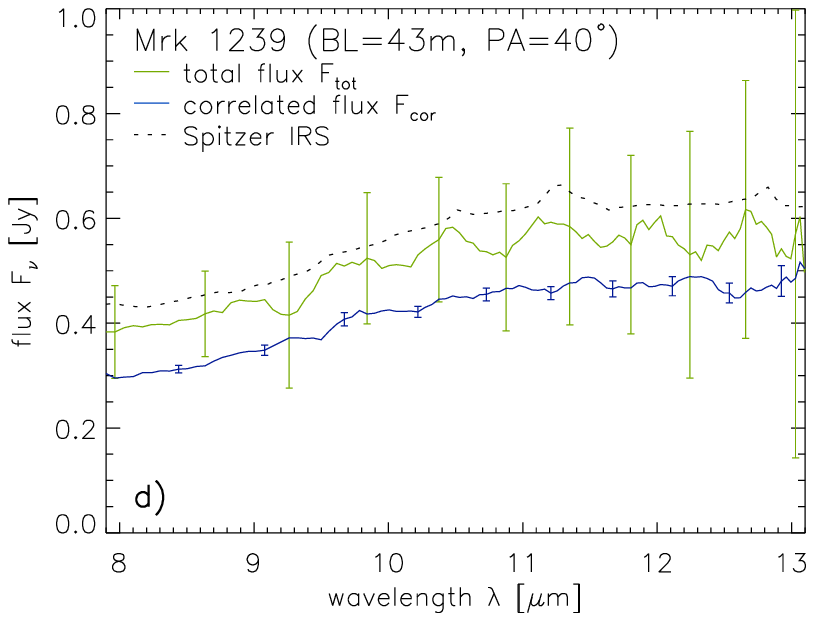}

\includegraphics{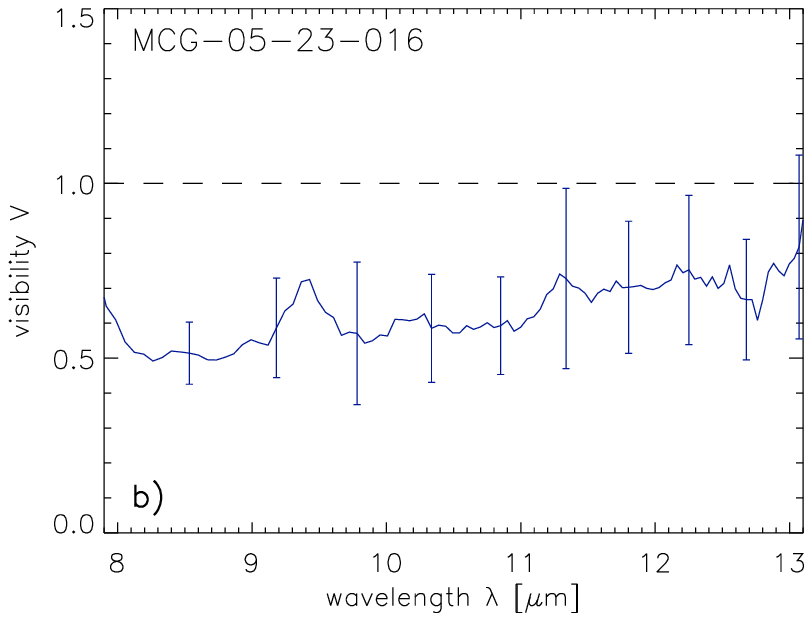}\hfill\includegraphics{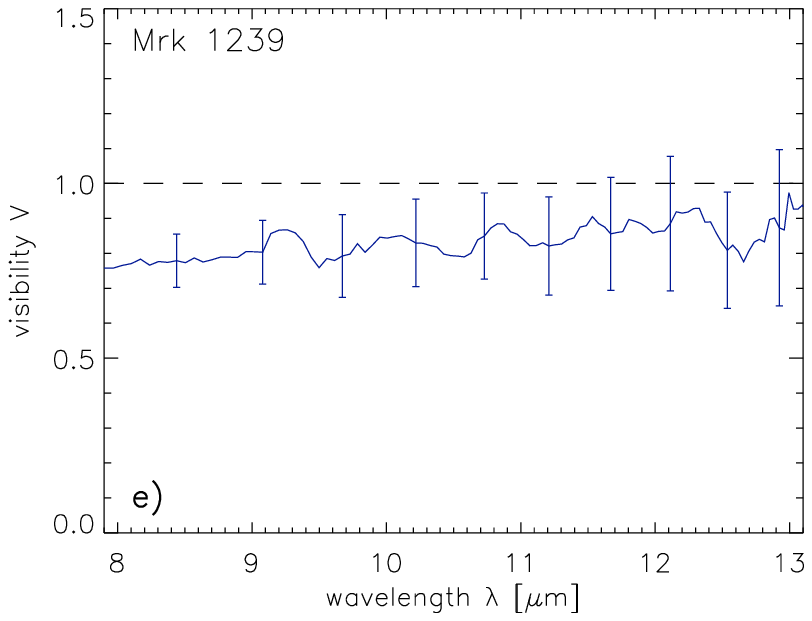}

\includegraphics{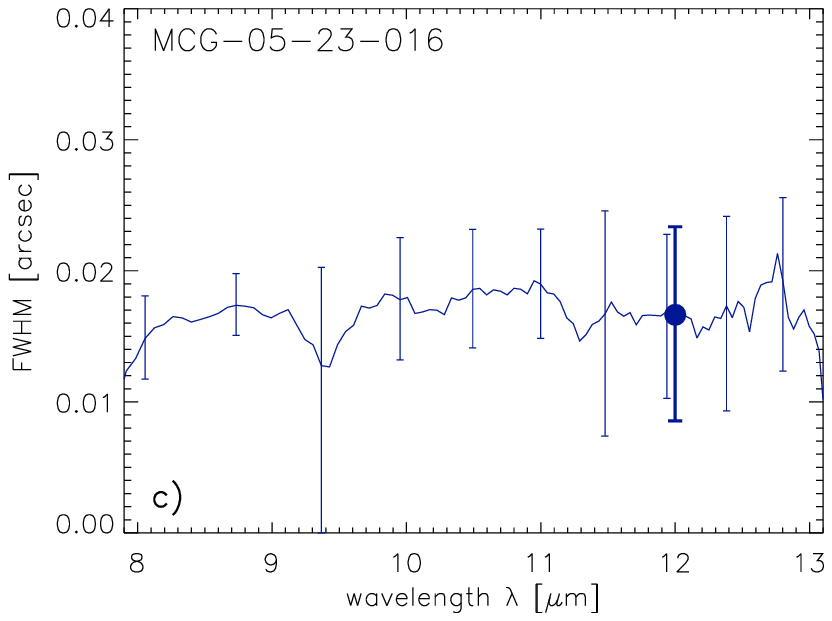}\hfill\includegraphics{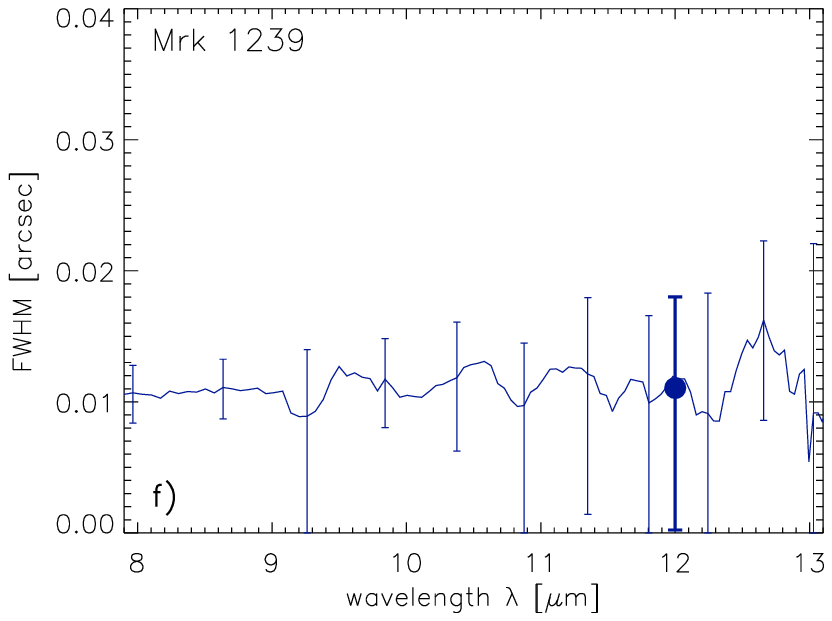}

\caption{Results of the interferometric measurements with MIDI for MCG-05-23-016
(\textit{left column}) and for Mrk~1239 (\textit{right column}).
In the first row, the correlated (blue) and the total (green) flux
spectra are displayed. For comparison, the VISIR spectrum for MCG-05-23-016
from \citet{2009Hoenig1} and the Spitzer spectrum for Mrk~1239 are
also plotted (dotted lines). In the second row, the visibilities are
shown. The last row shows the $\mathit{FWHM}$ of a corresponding
Gaussian flux distribution. The size estimates at $12\,\mathrm{\mu m}$
(see Sect.~\ref{sub:discussion_sizes}) are indicated by filled circles.
For the sake of clarity, the errors of the MIDI data are only indicated
for every tenth wavelength bin. }
\label{fig:result_mcg-and-mrk}
\end{figure*}
\afterpage{\clearpage}

The MIDI data are presented in Fig.~\ref{fig:result_ngc1365}. In
the left column, the correlated and total flux spectra for the three
different baseline lengths and orientations at which the source was
observed are shown together with the TIMMI2 spectrum from \citet{2004Siebenmorgen}.
Neither the total nor the correlated fluxes show indication of a silicate
feature, consistent with the featureless TIMMI2 spectrum. Our total
flux spectra show a continuous increase of the spectrum from $\mathrm{\sim}0.25\,\mathrm{Jy}$
at $8\,\mathrm{\mu m}$ to $\mathrm{\sim}0.50\,\mathrm{Jy}$ at $13\,\mathrm{\mu m}$,
except for the second measurement (Fig.~\ref{fig:result_ngc1365}b),
where up to $30\,\%$ lower flux values were measured. The MIDI spectra
thus seem to be slightly lower than the spectrum obtained with TIMMI2.
Nevertheless, all of the MIDI total flux spectra are consistent with
each other and with the TIMMI2 spectrum when taking into account their
large uncertainties. These uncertainties are caused by the residuals
of the background subtraction which are especially strong for the
second measurement. We consider the variations among our measurements
as an indication of the applicability of our data reduction methods
and of the true uncertainty in the determination of the total flux.
The correlated fluxes on the other hand rise from $\mathrm{\sim}0.15\,\mathrm{Jy}$
at $8\,\mathrm{\mu m}$ to $\mathrm{\sim}0.30\,\mathrm{Jy}$ at $13\,\mathrm{\mu m}$
and are hence below the level of the total flux measurements ($F_{\mathrm{cor}}<F_{\mathrm{tot}}$).
About $30$ to $40\,\%$ of the nuclear mid-infrared flux is resolved
by our interferometric set-up.

A comparison of the three correlated flux spectra and the corresponding
visibility spectra is given in Fig.~\ref{fig:result_ngc1365}d and
e respectively. Taking into account that a longer baseline length
means a higher spatial resolution, the source appears more extended
along $\mathit{PA}=31^{\circ}$ than along $\mathit{PA}=112^{\circ}$:
the correlated flux at $\mathit{PA}=31^{\circ}$ tends to be slightly
lower than that at $\mathit{PA}=112^{\circ}$, despite the longer
baseline and the consequently higher spatial resolution for the measurement
at $\mathit{PA}=112^{\circ}$. The same conclusion follows from the
visibilities: the two visibilities at $\mathit{PA}=93^{\circ}$ and
$112^{\circ}$ are still consistent with an unresolved source (i.e.
with $V=1$), while the visibility at $\mathit{PA}=31^{\circ}$ indicates
that the source is partially resolved.

The angular sizes derived from the visibilities according to Eq.~\ref{eq:faint_gauss-fwhm}
are displayed in Fig.~\ref{fig:result_ngc1365}f. The assumed Gaussian
brightness distribution has an angular extent between $5$ and $30\,\mathrm{mas}$.
Like the correlated fluxes or the visibilities, the size estimates
show evidence for an asymmetry with a larger extent of the source
($\mathrm{\sim}20\,\mathrm{mas}$) along $\mathit{PA}=31^{\circ}$
and a smaller size ($\mathrm{\sim}10\,\mathrm{mas}$) along $\mathit{PA}=112^{\circ}$.
At the distance of NGC~1365, $20\,\mathrm{mas}$ correspond to $2\,\mathrm{pc}$,
which sets a strong limit on the global size of the distribution of
warm dust in the nucleus of this galaxy. We note that, if our tentative
detection of an elongated structure along $\mathit{PA}\sim31^{\circ}$
can be confirmed, it would not agree with the E-W elongation claimed
by \citet{2005Galliano}. It would, however, lie more or less perpendicular
to the ionisation cone at $\mathit{PA}\sim130^{\circ}$ \citep{1994Wilson,1996Hjelm}
and the radio {}``jet'' at $\mathit{PA}=125^{\circ}$ \citep{2000Kinney}.

\subsection{MCG-05-23-016\label{sub:result_mcg0523016}}

MCG-05-23-016 (ESO~434-G40) is a lenticular Seyfert 2 galaxy at a
distance of $35\,\mathrm{Mpc}$ ($v_{\mathrm{hel}}=2544\pm12\,\mathrm{km}\,\mathrm{s}^{-1}$,
\citealp{2003Wegner}; $1\,\mathrm{arcsec}=170\,\mathrm{pc}$), which
has been studied mostly for its X-ray emission (e.g. \citealp{2004Balestra,2004Mattson}).
Its mid-infrared spectrum is characteristic of this type of nucleus,
rising strongly towards longer wavelengths and it displays a very
shallow silicate absorption feature (see also \citealp{1982Frogel}).

The data obtained with MIDI are shown in the left column of Fig.~\ref{fig:result_mcg-and-mrk}.
The values measured for the total flux are on the order of $0.3\,\mathrm{Jy}$
from $8$ to $10\,\mu\mathrm{m}$ and then continuously increase to
$0.8\,\mathrm{Jy}$ at $13\,\mu\mathrm{m}$. Hence, the spectrum is
in excellent agreement with the VISIR spectrum presented in \citet{2009Hoenig1}
and with our TIMMI2 photometry: $0.65\pm0.10\,\mathrm{Jy}$ at $11.9\,\mu\mathrm{m}$.
We are confident that the total flux values measured by MIDI are accurate
for this galaxy. The correlated flux is considerably lower than the
total flux ($F_{\mathrm{cor}}<F_{\mathrm{tot}}$), indicating that
the emitter in the mid-infrared is partially resolved: roughly between
$30\,\%$ (at $13\,\mu\mathrm{m}$) and $50\,\%$ (at $8\,\mu\mathrm{m}$)
of the nuclear flux are resolved out by our interferometric measurement.
This is also reflected by the visibility which is on the order of
$50$ to $70\,\%$ (see Fig.~\ref{fig:result_mcg-and-mrk}b).

\begin{figure}
\centering

\includegraphics{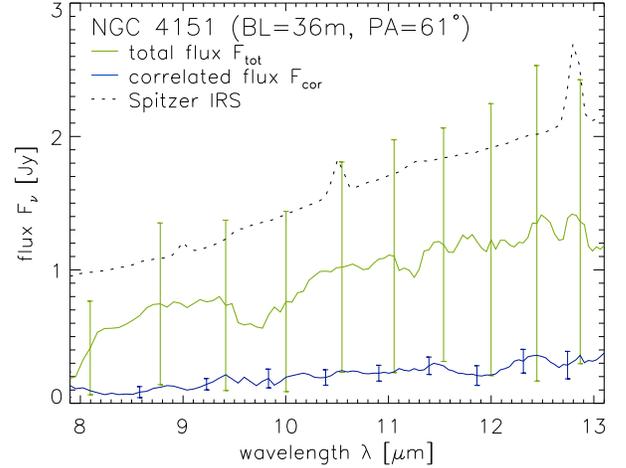}

\caption{Correlated (blue) and total (green) flux spectra for the nucleus of
NGC~4151. Also plotted is the spectrum obtained with Spitzer (dotted
line) which clearly shows the {[}\ion{Ar}{iii}], {[}\ion{S}{iv}]
and {[}\ion{Ne}{ii}] emission lines at $9.0\,\mathrm{\mu m}$, $10.5\,\mathrm{\mu m}$
and $12.8\,\mathrm{\mu m}$ respectively (cf. also Fig.~1 in \citealt{2005Weedman}).
For the sake of clarity, the errors of the MIDI data are only indicated
every tenth wavelength bin. }
\label{fig:result_ngc4151}
\end{figure}

Assuming a Gaussian flux distribution, an upper limit for the emission
region in MCG-05-23-016 can be derived: at $\lambda=12\,\mu\mathrm{m}$
it has a size of less than $23\,\mathrm{mas}$, which corresponds
to less than $4.0\,\mathrm{pc}$ at the distance of this galaxy (see
thick error bar at $\lambda=12\,\mu\mathrm{m}$ in Fig.~\ref{fig:result_mcg-and-mrk}b).
From the fact that this source is slightly resolved, we similarly
deduce a lower limit for the size of the emission region, which is
roughly $9\,\mathrm{mas}$ and corresponds to $1.5\,\mathrm{pc}$
in MCG-05-23-016.

\subsection{Mrk~1239\label{sub:result_mrk1239}}

Mrk~1239 is a narrow-line Seyfert 1 (NLS1) galaxy located at a distance
of about $80\,\mathrm{Mpc}$ ($v_{\mathrm{hel}}=5747\,\mathrm{km}\,\mathrm{s}^{-1}$,
\citealp{1995Beers}; $1\,\mathrm{arcsec}=380\,\mathrm{pc}$). Its
unresolved near-infrared emission shows a strong bump at $2.2\,\mathrm{\mu m}$
which is interpreted as very hot dust near its sublimation temperature
($T\sim1200\,\mathrm{K}$), very likely located both in the upper
layers of the torus and close to the apex of the polar scattering
region \citep{2006Rodriguez}. While the source appears unresolved
with a flux of $0.64\pm0.10\,\mathrm{Jy}$ in the TIMMI2 image at
$11.9\,\mu\mathrm{m}$ \citep{2008Raban}, \citet{2004Gorjian} detected
an approximately $1\,\mathrm{arcsec}$ long extension from the nucleus
to the northwest.

The correlated and total flux spectra, the visibility spectrum as
well as the derived sizes for a Gaussian brightness distribution are
displayed in the right column of Fig.~\ref{fig:result_mcg-and-mrk},
panels d)--f), respectively. The MIDI total flux spectrum is rising
from $0.4$ to $0.6\,\mathrm{Jy}$. It is slightly lower but still
consistent with the Spitzer spectrum as well as with the $0.64\,\mathrm{Jy}$
measured with TIMMI2 at $11.9\,\mu\mathrm{m}$ (cf. Table~\ref{table:target-list-original}).
The overall shapes of the MIDI and Spitzer spectra agree well: they
both exhibit the silicate feature in emission (slightly convex shape
of the spectra). The good agreement between the MIDI and Spitzer spectra
despite the difference in apertures can be explained by the nature
of Mrk~1239 as a Seyfert 1 galaxy without evidence of a starburst,
thus implying a compact, solely AGN dominated emission region in the
mid-infrared. The correlated flux spectrum is slightly lower than
the total flux spectrum, although they are consistent with each other
within the errors. As a consequence, the mid-infrared source in this
nucleus is only slightly resolved, if at all. This can also be seen
from the visibility, which is larger than $80\,\%$, with errors on
the order of $20\,\%$. From the upper limit on the full width at
half maximum of an assumed Gaussian flux distribution we conclude
that the mid-infrared source in Mrk~1239 most likely has a size of
less than $\mathrm{\sim}18\,\mathrm{mas}$, that is, less than $7\,\mathrm{pc}$.

\begin{figure*}
\centering

\includegraphics{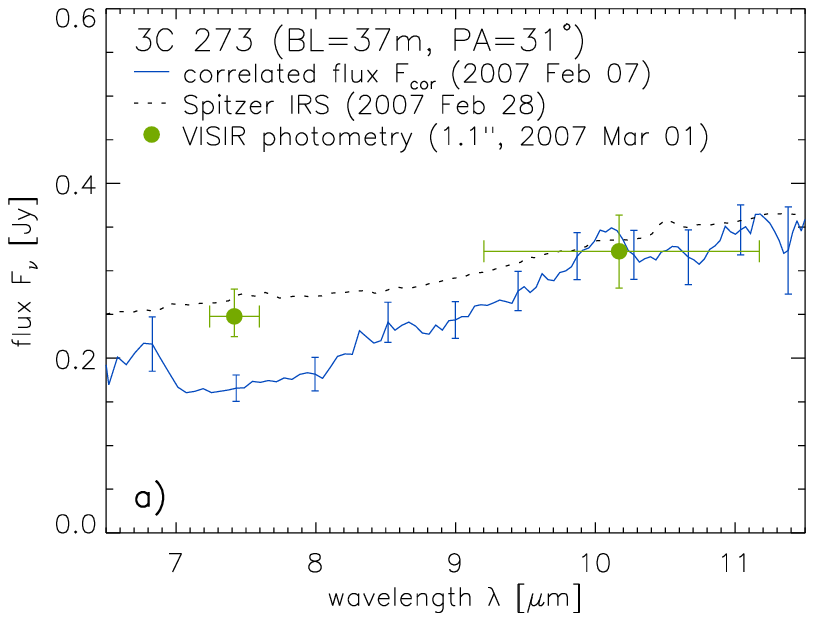}\hfill{}\includegraphics{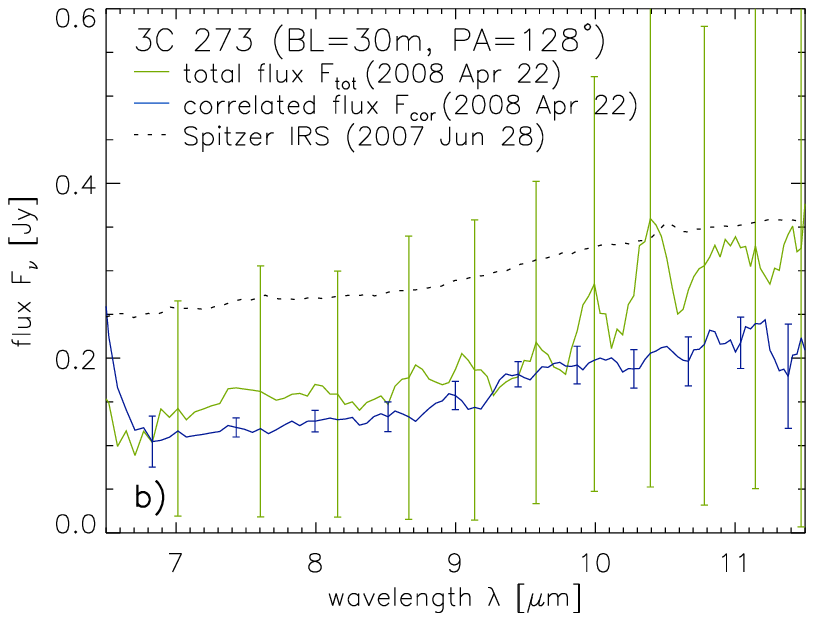}

\caption{Results for 3C~273. Panel \textbf{a)} shows the spectrum of the correlated
flux measured with MIDI on 2007 Feb 07 (blue line) compared to the
intermediate band photometry with VISIR (green dots) and a Spitzer
spectrum from 2007 Feb 28 (dotted line). For the VISIR data, the error
bars in the wavelength direction correspond to the width of the filter
at half of the maximum transmission. In panel \textbf{b)}, the MIDI
measurements from 2008 Apr 22 are plotted together with a Spitzer
spectrum from 2007 Jun 28. The Spitzer spectrum in panel \textbf{b)}
was observed four months after the one displayed in panel \textbf{a)},
nevertheless both are almost identical. Due to the relatively high
redshift of this source ($z=0.158$), the wavelength range plotted
was shifted. For the sake of clarity, the errors of the MIDI data
are only indicated every tenth wavelength bin. }

\label{fig:result_3c273}
\end{figure*}

\subsection{NGC~4151\label{sub:result_ngc4151}}

At a distance of $13.6\,\mathrm{Mpc}$ ($v_{\mathrm{hel}}=997\pm3\,\mathrm{km}\,\mathrm{s}^{-1}$,
\citealp{1992Pedlar}; $1\,\mathrm{arcsec}=65\,\mathrm{pc}$), NGC~4151
hosts the closest and brightest ($F_{\mathrm{N}}=1.4\,\mathrm{Jy}$,
\citealt{2003Radomski}) Seyfert~1 nucleus. It has been extensively
studied at all wavelengths. High resolution mid-infrared observations
are presented in \citet{1990Neugebauer}, \citet{2003Radomski} and
\citet{2003Soifer}. They show that $\mathrm{\leq}70\,\%$ of the
total mid-infrared flux comes from a core component with a size of
$\mathrm{\lesssim}10\,\mathrm{pc}$, while the rest originates in
extended emission from dust in the narrow line region. Observations
with the Keck Interferometer have shown that the near-infrared emission
of the nucleus is very compact ($\mathrm{\leq}0.1\,\mathrm{pc}$,
\citealt{2003Swain}).

The results of the MIDI observations of NGC~4151 from February 2007
are displayed in Fig.~\ref{fig:result_ngc4151} together with the
Spitzer spectrum. The total flux spectrum rises from $\mathrm{\sim}0.5\,\mathrm{Jy}$
at $8\,\mathrm{\mu m}$ to $\mathrm{\sim}1.3\,\mathrm{Jy}$ at $13\,\mathrm{\mu m}$.
Its general shape agrees with the shape of the Spitzer spectrum, although
it is lower by a factor of $30$ to $50\,\%$. None of the smaller
scale variations can be trusted and the decrease of flux below $8.2$
and around $9.7\,\mathrm{\mu m}$ is due to an inadequate removal
of the absorption caused by water and ozone in the Earth's atmosphere
during calibration, especially at the high airmass at which this source
was observed. The Spitzer and the MIDI spectra are consistent with
the flux being distributed in an unresolved {}``core'' and an extended
component as described by \citet{2003Radomski} when accounting for
the difference in the apertures. The correlated flux spectrum is even
lower, on a level between $0.1$ and $0.3\,\mathrm{Jy}$, implying
that the mid-infrared source is clearly resolved. With a baseline
length of $\mathit{BL}=35.8\,\mathrm{m}$ and assuming a visibility
of $V=F_{\mathrm{cor}}/F_{\mathrm{tot}}\sim20\,\%$ at $\lambda\sim12\,\mu\mathrm{m}$,
we obtain a size of the emission region on the order of $45\,\mathrm{mas}$,
which is $\mathrm{\sim}3\,\mathrm{pc}$ at the distance of NGC~4151.
That the emission appears to be resolved on such scales shows that
most of the mid-infrared emission is of thermal origin. The size derived
here is significantly smaller than the $10\,\mathrm{pc}$ from \citet{1990Neugebauer}
but well above the estimated inner radius of the torus, $r_{\mathrm{inner}}=0.04\,\mathrm{pc}$
\citep{2004Minezaki}. A full analysis of the MIDI data of this source
will be given in \citealt{2009Burtscher}.

\subsection{3C~273\label{sub:result_3c273}}

3C~273 (PG 1226+023) was the first quasar to be discovered \citep{1963Schmidt}.
As it is the brightest object of its kind and because it exhibits
a one-sided jet which is visible from the radio regime to $\gamma$-rays,
it is one of the best observed and studied AGN. The quasar is located
at a redshift of $z=0.158$, which corresponds to a distance of $650\,$$\mathrm{Mpc}$
($1\,\mathrm{arcsec}=3.1\,\mathrm{kpc}$, \citealp{1963Schmidt}).
A detailed review of the source is given by \citet{1998Courvoisier}.

The results of our measurements are shown in Fig.~\ref{fig:result_3c273}.
Due to the relatively high redshift, the restframe wavelengths probed
by MIDI are between $6.5$ and $11.5\,\mu\mathrm{m}$. For the first
measurement (panel a)), no total flux measurement could be obtained;
only intermediate band photometry could be carried out with VISIR
21 days after the interferometric observations (see Appendix~\ref{sub:observ_3c273}).
For the second measurement (panel b)), the total flux spectrum has
very large errors. Therefore the interferometric data are compared
to Spitzer spectra. These were obtained on two different dates: 2007
Feb 28 (one day before the observations with VISIR) and 2007 Jun 28.
Both Spitzer spectra are almost identical. They are well suited for
comparison to the MIDI data because of the pointlike nature of the
nucleus in 3C~273: neither in our VISIR imaging (cf. Appendix~\ref{sub:observ_3c273})
nor in our observations with TIMMI2 \citep{2008Raban} are there signs
for extended emission. Furthermore, the VISIR measurements are consistent
with the Spitzer spectra, as are the TIMMI2 measurements. We interpret
this as an indication that the total flux measured by MIDI in 2008
underestimates the true flux of the source.

\begin{figure}
\centering

\includegraphics{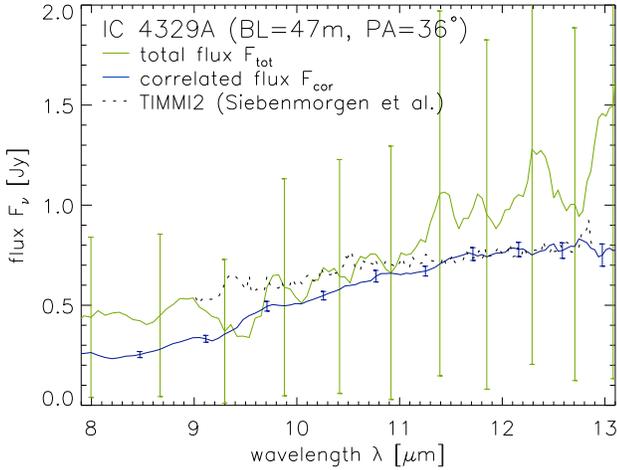}
\caption{Results of the MIDI measurements for IC~4329A. The correlated flux
spectrum (blue) and the total flux spectrum (green) are compared to
the TIMMI2 spectrum from \citet[dotted line]{2004Siebenmorgen}. For
the sake of clarity, the errors of the MIDI data are only indicated
every tenth wavelength bin. }
\label{fig:result_ic4329a}
\end{figure}

The correlated fluxes both rise between $6.5$ and $11.5\,\mu\mathrm{m}$.
This is mainly due to the silicate feature in emission, which has
its peak at $11\,\mu\mathrm{m}$ when looking at the entire mid-infrared
spectrum (cf. Fig.~1 in \citealt{2005Hao}). The measurement on 2007
Feb 07 has a relatively steep, increasing spectrum and flux values
consistent with the Spitzer spectrum at the long wavelength end. The
source appears to be only resolved to a minor fraction and only at
shorter wavelengths. The correlated fluxes from 2008 Apr 22, on the
other hand, follow the Spitzer spectrum at levels of $50$ to $60\,\%$.
It thus seems that up to half of the flux of the source is resolved,
and this despite the shorter baseline length and thus smaller spatial
resolution. The difference between the correlated fluxes may have
two causes: (1) the difference in baselines or (2) the variability
of the source. The two interferometric measurements were taken with
almost perpendicular baseline orientations ($\mathit{PA}\sim31^{\circ}$
and $\mathit{PA}\sim128^{\circ}$). Changes in the correlated flux
could thus be interpreted as an indication for an elongated morphology.
On the other hand, variability of up to $\mathrm{\sim}60\,\%$ has
been observed in the mid-infrared at $10.6\,\mathrm{\mu m}$ \citep{1999Neugebauer}
for 3C~273, so that the variations could have been caused by intrinsic
changes in the spectrum of the quasar. All Spitzer spectra (the two
Spitzer spectra shown in Fig.~\ref{fig:result_3c273} from 2007 Feb
28 and 2007 Jun 28 as well as the spectrum in \citealt{2005Hao} from
2004 Jan 06) show variations of less than $10\,\%$ and are also consistent
with the VISIR and TIMMI2 photometries. Hence we see no evidence for
any significant variability during our observations. This is also
supported by submm data of the source obtained as part of the monitoring
program initiated by the SMA \citep{2007Gurwell}: 3C~273 appeared
in a relatively quiescent state in February 2007 and April 2008. This
leads to the conclusion that we mainly measured the emission from
warm dust which is not expected to change on shorter timescales and
that the change in the correlated flux is more likely to be caused
by an elongated morphology. A more detailed analysis of the MIDI
data for 3C~273 will be given in \citealt{2009Jaffe}.

\begin{figure}
\centering

\includegraphics{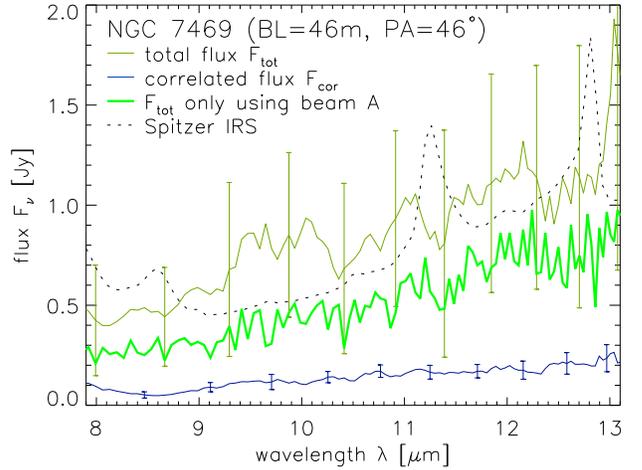}

\caption{Correlated (blue) and total (green) flux spectra of NGC~7469. For
comparison, the flux measured by Spitzer is traced by the dotted line
(cf. also Fig.~6 in \citealt{2005Weedman}). The total flux, only
determined using beam A (UT3), is plotted in dark green. For the sake
of clarity, error bars are only given every tenth wavelength bin. }
\label{fig:result_ngc7469}
\end{figure}

\subsection{IC~4329A\label{sub:result_ic4329a}}

IC~4329A is an edge-on spiral galaxy at a distance of $65\,\mathrm{Mpc}$
($4793\pm27\,\mathrm{km}\,\mathrm{s}^{-1}$, $1\,\mathrm{arcsec}=320\,\mathrm{pc}$,
\citealt{1991deVaucouleurs}) hosting a Seyfert 1.2 nucleus. The nuclear
activity might be triggered by the proximity to IC 4329, a giant elliptical
galaxy. As one of the brightest X-ray sources, it has been mainly
studied in this wavelength range.

Fig.~\ref{fig:result_ic4329a} shows the results of the MIDI measurements
as well as the spectrum obtained with TIMMI2 from \citet{2004Siebenmorgen}.
While the MIDI correlated flux spectrum seems to be well determined,
the total flux spectrum has large uncertainties, mainly because the
measured fluxes for the two telescopes differ by more than a factor
of 2 (cf. Appendix~\ref{sub:observ_ic4329a}). The total flux spectrum
measured by MIDI lies between the slightly lower TIMMI2 spectrum and
the slightly higher spectra obtained with larger apertures, e.g. the
Spitzer IRS spectrum (not shown here) or the spectrum presented in
\citet{1991Roche}. As the total flux measurement with MIDI is consistent
with these other spectra, we assume the total flux of the source to
rise from $\mathrm{\sim}0.5\,\mathrm{Jy}$ at $8\,\mu\mathrm{m}$
to $1.1\,\mathrm{Jy}$ at $13\,\mu\mathrm{m}$ for the following.
The correlated flux is slightly lower but nevertheless consistent
with these values considering the uncertainties. The conclusion is
that the nucleus of IC~4329A is essentially unresolved. This means
that the bulk of the mid-infrared emission is concentrated on scales
smaller than the spatial frequency corresponding to the baseline length
for our measurement ($\mathit{BL}=46.6\,\mathrm{m}$), that is, on
scales roughly smaller than $30\,\mathrm{mas}$, which corresponds
to $\mathrm{\sim}10\,\mathrm{pc}$ in IC~4329A.

\subsection{NGC~7469\label{sub:result_ngc7469}}

NGC~7469 is a well-studied, barred spiral galaxy at a distance of
about $65\,\mathrm{Mpc}$ ($v_{\mathrm{hel}}=4843\pm5\,\mathrm{km}\,\mathrm{s}^{-1}$,
\citealt{2002Beswick}; $1\,\mathrm{arcsec}=320\,\mathrm{pc}$),
which hosts a classical Seyfert~1 nucleus. It is also classified
as a Luminous InfraRed Galaxy (LIRG) and it harbours a face-on circumnuclear
starforming ring with a size of $1.0\,\mathrm{kpc}$ ($3\,\mathrm{arcsec}$),
which has been detected at radio, optical and infrared wavelengths
\citep[and references therein]{2007Diaz-Santos}. High resolution
studies in the mid-infrared have been carried out by \citet{1994Miles,2003Soifer,2004Gorjian}
and \citet{2005Galliano}. They have found that $50$ to $60\,\%$
of the total mid-infrared emission in the centre of this galaxy (i.e.
within $\mathrm{\sim}6\,\mathrm{arcsec}=2\,\mathrm{kpc}$) originate
from the star forming ring, while only the remaining $40\,\%$ to
$50\,\%$ ($\mathrm{\sim}0.5\,\mathrm{Jy}$) can be ascribed to the
compact nuclear source. Using deconvolved images obtained with the
Keck telescope, \citet{2003Soifer} marginally resolve the nuclear
source. They report a structure of $(\mathrm{<}0.04)\times0.08\,\mathrm{arcsec}$
($13\times26\,\mathrm{pc}$) with a position angle of $135^{\circ}$.

The results of the first MIDI observing run are shown in Fig.~\ref{fig:result_ngc7469}.
These, however, need to be discussed: we find a significant difference
between the pipeline reduced total flux spectrum (thin green line)
and the one obtained only from beam A (UT3, see Appendix~\ref{sub:observ_ngc7469};
thick green line). Which one is credible? The best mid-infrared spectrum
available for comparison is from Spitzer. To a large degree, it contains
the flux from the star forming ring, as can be deduced from the prominent
PAH features and emission lines. It is therefore an extremely conservative
upper limit for the MIDI data and the expectation from the mid-infrared
imaging is that the total flux measured by MIDI should be roughly
half of that of Spitzer. As the pipeline reduced spectrum lies above
that of Spitzer for most of the N band, we consider this determination
to be erroneous. On the other hand, the total flux spectrum only determined
using beam A is on levels of roughly $50$ to $80\,\%$ of the continuum
component in the Spitzer spectrum and it is consistent with the photometric
values published for the nuclear component in \citet{2003Soifer,2005Galliano}
and \citet{2008Raban}: $0.65\,\mathrm{Jy}$ at $12.5\,\mathrm{\mu m}$,
$0.56\,\mathrm{Jy}$ at $11.9\,\mathrm{\mu m}$ and $0.67\,\mathrm{Jy}$
at $11.9\,\mathrm{\mu m}$ respectively. Therefore we consider the
total flux only using the A photometry as the best estimate for the
total flux spectrum, $F_{\mathrm{tot}}$.

From Fig.~\ref{fig:result_ngc7469} it appears that the correlated
flux spectrum, with $F_{\mathrm{cor}}\sim0.1-0.2\,\mathrm{Jy}$, is
significantly lower than the total flux spectrum, i.e. $F_{\mathrm{cor}}<F_{\mathrm{tot}}$.
Our interferometric observations were obtained with a position angle
of $\mathit{PA}=45^{\circ}$, which is more or less along the minor
axis of the extended structure from \citet{2003Soifer}. The fringe
spacing of the interferometer with a baseline length of $\mathit{BL}=46\,\mathrm{m}$
is on the order of $25\,\mathrm{mas}$ ($\mathrm{=}8\,\mathrm{pc}$
in NGC~7469). Due to the large errors in our data and especially
the large uncertainty in the total flux spectrum, the determination
of a size estimate is very ambitious. But considering that fringes
could be tracked and that the correlated flux seems to be significantly
lower than the total flux, the overall size of the source cannot be
much different from the fringe spacing. Hence, we estimate the size
of the dust distribution to be on the order of $10\,\mathrm{pc}$.
Note that, for the measurement on the second baseline with $\mathit{BL}=51\,\mathrm{m}$
and $\mathit{PA}=107^{\circ}$, the correlated flux seems to have
been even lower, because no fringes could be tracked (see Appendix~\ref{sub:observ_ngc7469}).
This could be an indication that the source is slightly more resolved
in this direction. This would be in rough agreement with the extended
structure described by \citet{2003Soifer}. More data with a higher
signal to noise ratio will be needed to confirm this speculation.

\section{Summary of the observations and their implications\label{sec:summary_observations}}

In the MIDI AGN snapshot survey, interferometric observations in the
mid-infrared were attempted for a set of ten AGN. Out of these ten
sources, an interferometric signal could be detected for eight sources
and only for two no measurement with MIDI was possible at all. For
the latter two sources (NGC~3281 and NGC~5506, see Appendices \ref{sub:observ_ngc3281}
and \ref{sub:observ_5506}), no stable AO correction could be obtained
with MACAO using the optical nuclei of these two galaxies. Hence,
the observation failed during the acquisition process and not because
an interferometric signal could not be detected by MIDI. For all of
the eight sources where a stable AO correction could be achieved at
some point, an interferometric signal was detected. This means that
the true detection rate of a fringe signal by MIDI itself for the
AGN in the snapshot survey is $100\,\%$. For all except one of these
eight sources, fringes could also be tracked, that is, the zero OPD
position could be continuously determined. Only for IRAS~05189-2524,
the online tracking software was not able to determine the zero OPD
position. This was only possible in post-processing (see Appendix~\ref{sub:observ_leda17155}).
For six of the seven remaining sources, the faintness and extent of
the nuclei in the optical were still challenging for MACAO. Mainly
the photometric observations were affected by having to use the AO
systems at their limits: in many cases a stable AO correction was
not possible while chopping. The chopping constitutes an additional
complication for the AO system, since in every chopping cycle the
AO loop has to be opened for the off target position (sky position)
and then closed again for the on target position. On the other hand,
for the interferometric measurement no chopping is needed and generally
a more stable AO correction is achieved (see Sect.~\ref{sec:observations}).
Only for MCG-05-23-016 were there no AO problems at all. In fact,
this is the only galaxy from the snapshot survey, where the AO correction
was performed using a separate Coudé guide star. For all other galaxies,
the more or less extended and obscured nuclei were used.

The overall perfomance of the adaptive optics systems depends on several
factors, the most important being (1) the performance of the individual
adaptive optics units, (2) the brightness and spatial extent of the
object used for guiding and (3) the atmospheric conditions. For (1)
we note that, during our observations, the MACAO unit on UT2 caused
the bulk of the problems. In general, this unit produced a worse wavefront
correction -- resulting in a reduced PSF quality -- than the other
units and it was often this unit which failed totally to close the
loop. A more robust correction is achieved using bright and pointlike
Coudé guiding sources. To date, for all except one of the faint sources
observations have been tried with Coudé guiding on the nucleus, although
for some of the sources a possible guide star is in range (see Table~\ref{table:target-list-observations}).
Especially in the case of NGC~7469, and maybe also for NGC~3281,
future observations should be tried using the guide star available.
Finally, the probability of a stable AO correction depends sensitively
on the atmospheric conditions at the time of observation: good seeing
conditions and long coherence times increase the chances of a stable
AO correction and hence are necessary for the interferometric observation
of faint AGN with the VLTI.

From the high detection rate of fringes in the snapshot survey, we
deduce that the ultimate limit for the detection of an interferometric
signal with MIDI has not yet been reached. From our data on IRAS~05189-2524,
we infer that the limit for the detection and tracking of fringes
with MIDI is on the order of $0.1$ to $0.2\,\mathrm{Jy}$. This limit
is however only reached when an optimised mask and optimised data
reduction settings are applied. At the moment, this can only be done
in post processing. For the current implementation of the software
at the VLTI, the limit is approximately a factor of 2 higher, on the
order of $0.3\,\mathrm{Jy}$. Simply using longer integration times
and redundant measurements will not improve the detectability of even
weaker sources, as it will not be possible to continuously determine
the zero OPD position in real time. An external fringe tracker will
be necessary, if even fainter sources are to be observed. Furthermore,
we note that the detection and the recording of the interferometric
signal is normally not the only goal of a MIDI observation. Instead,
full data sets consisting of both the interferometric signal as well
as the photometry are required to scientifically evaluate the data.
This is especially the case if only a few visibility points are observed
(as is the case for the sources presented here) and if a direct comparison
between several measurements of the correlated flux at different baseline
lengths and orientations is not possible. Even if a stable AO correction
is achieved, the main limitations are still connected with the robust
determination of the total flux spectrum from the photometry: it is
the total fluxes that show large (statistical) errors due to the constant
variability of the background in the mid-infrared. Even for stable
atmospheric conditions, total fluxes on the order of $1.0\,\mathrm{Jy}$
have a statistical error of up to $20\,\%$. Less stable atmospheric
conditions immediately lead to larger statistical errors, as can be
directly deduced from the MIDI observations of Centaurus~A \citep[see Fig. 1 in][]{2007Meisenheimer}.
Due to these limitations, for very faint sources, it may be better
to skip the photometric measurements and use relative changes among
the correlated fluxes to characterise the emitter. This is, in fact,
how observations are commonly carried out with radio interferometry.

In general, the MIDI data nevertheless appear to be reasonable and
trustworthy. Especially the good agreement between the MIDI measurements
and the mid-infrared data used for comparison in the case of MCG-05-23-016
and Mrk~1239 is reassuring. It shows that MIDI produces reliable
results even at its sensitivity limit when the observational strategy
is adapted accordingly and when complete data sets are obtained. For
faint sources, it is imperative to carry out MIDI observations with
significantly increased exposure times (e.g. $\mathrm{NDIT}=8000$
instead of $\mathrm{NDIT}=2000$ for the photometry) and a certain
degree of redundancy (e.g. by repeating the error-prone photometric
observations) in order to reach an acceptable signal-to-noise ratio.
This and the often longer acquisition procedure lead to a considerably
larger amount of time, up to $1\,\mathrm{h}$, needed for the measurement
of the science target alone. Together with the calibrator observations,
a full, calibrated visibility point for a faint source thus requires
up to $1.5\,\mathrm{h}$ of observing time.

\section{Discussion\label{sec:discussion}}

Together with the other AGN observed with MIDI, the data presented
here make up the spectra with the highest spatial resolution of cores
of AGN in the mid-infrared. They allow us to directly analyse the
properties of the circumnuclear dust which is heated by the central
engine of the AGN.

\begin{table}
\caption{Characteristics derived from the interferometric observations of the
AGN in the snapshot survey (marked by asterisks) and other AGN studied
by MIDI.\label{table:discussion_characteristics}}

\centering

\begin{tabular}{@{ \extracolsep{1.0pt}}r@{ \extracolsep{1.0pt}}l@{ \extracolsep{3.0pt}}l@{ \extracolsep{3.0pt}}r@{ \extracolsep{3.0pt}}r@{ \extracolsep{3.0pt}}r@{ \extracolsep{3.0pt}}r@{ \extracolsep{3.0pt}}r}
\toprule 
\hline  & Galaxy &  & $D$ & $\mathit{BL}$ & $F_{\mathrm{tot},12\mathrm{\mu m}}$ & $F_{\mathrm{cor,12\mathrm{\mu m}}}$ & $s_{12\mathrm{\mu m}}$\tabularnewline
 & name & type & {[}Mpc] & {[}m] & {[}Jy] & {[}Jy] & {[}pc]\tabularnewline
 & (1) & (2) & (3) & (4) & (5) & (6) & (7)\tabularnewline
\midrule
 & NGC~1068 & Sy~2 & 14 & 33.6 & 16.50 & 4.00 & 3.2\tabularnewline
 & NGC~1068 & Sy~2 & 14 & 79.9 & 16.50 & 1.20 & \textgreater{}1.8\tabularnewline
{*} & NGC~1365 & Sy~1.8 & 18 & 46.6 & 0.51 & 0.28 & 2.0\tabularnewline
{*} & NGC~1365 & Sy~1.8 & 18 & 54.4 & 0.34 & 0.25 & \textless{}2.7\tabularnewline
{*} & NGC~1365 & Sy~1.8 & 18 & 62.3 & 0.47 & 0.32 & 1.1\tabularnewline
{*} & MCG-05-23-016 & Sy~2 & 35 & 46.1 & 0.60 & 0.35 & 2.8\tabularnewline
{*} & Mrk~1239 & Sy~1.5 & 80 & 43.0 & 0.57 & 0.48 & \textless{}7.0\tabularnewline
 & NGC~3783 & Sy~1 & 40 & 68.6 & 1.25 & 0.50 & 3.6\tabularnewline
 & NGC~3783 & Sy~1 & 40 & 64.9 & 1.02 & 0.54 & 3.1\tabularnewline
{*} & NGC~4151 & Sy~1.5 & 14 & 35.8 & 1.21 & 0.27 & \textgreater{}2.6\tabularnewline
{*} & 3C~273 & QSO & 650 & 36.7 & 0.32 & 0.32 & \textless{}67.\textcolor{white}{3}\tabularnewline
{*} & 3C~273 & QSO & 650 & 29.7 & 0.29 & 0.20 & \textless{}108.\textcolor{white}{7}\tabularnewline
{*} & IC~4329A & Sy~1.2 & 65 & 46.6 & 1.00 & 0.75 & \textless{}10.8\tabularnewline
 & Circinus & Sy~2 & 4 & 20.7 & 10.20 & 1.80 & \textgreater{}1.5\tabularnewline
 & Circinus & Sy~2 & 4 & 62.4 & 10.20 & 1.10 & \textgreater{}0.6\tabularnewline
{*} & NGC~7469 & Sy~1.2 & 65 & 46.4 & 0.70 & 0.17 & 10.5\tabularnewline
\bottomrule
\end{tabular}

~

\begin{minipage}[t][1\totalheight]{1\columnwidth}%
Notes: The columns are (1) the name, (2) the type and (3) the distance
of the galaxy (all repeated from Tab.~\ref{table:target-list-original});
(4) the baseline length for the respective observation; (5) the total
and (6) correlated fluxes at $12\,\mu\mathrm{m}$ as well as (7) the
approximate size of the emitter at $12\,\mu\mathrm{m}$.%
\end{minipage}%

\end{table}

In the following, two aspects of the sources will be discussed in
greater detail, namely the properties of their spectra (Sect.~\ref{sub:discussion_spectra})
and the size of their emission region (Sect.~\ref{sub:discussion_sizes}).
The size estimates used in the following discussion are those derived
in Sect.~\ref{sec:results}, taken at a wavelength of $12\,\mu\mathrm{m}$.
For those galaxies where plots of the wavelength dependency of the
size are given, i.e. for NGC~1365, MGC-05-23-016 and Mrk~1239, the
values are marked by filled circles in the respective figures (Figs.
\ref{fig:result_ngc1365} and \ref{fig:result_mcg-and-mrk}). 

The values of the sizes at $12\,\mu\mathrm{m}$ are listed in Table~\ref{table:discussion_characteristics}
for every visibility point together with the galaxy type, the distance
of the galaxy and the baseline length of the MIDI observation. Also
listed are the correlated and the total flux values at $12\,\mu\mathrm{m}$,
$F_{\mathrm{cor,12\mathrm{\mu m}}}$ and $F_{\mathrm{tot},12\mathrm{\mu m}}$.
For $\lambda=12\,\mu\mathrm{m}$, the MIDI measurements have the highest
signal to noise ratio and they are least affected by the edges of
the N band or the atmospheric ozone absorption feature.

The values for the sources of the snapshot survey were complemented
by values for selected visibility points of other Seyfert galaxies,
for which MIDI results have been published, that is, for NGC~1068
\citep{2009Raban}, for the Circinus galaxy \citep{2007Tristram2}
and for NGC~3783 \citep{2008Beckert}. The sizes for these additional
sources were determined in the same way as for the sources of the
snapshot survey. Due to its nature as a radio galaxy with a strong
non-thermal component in the mid-infrared, Centaurus~A was not included
in the current analysis.

\begin{figure*}
\centering

\includegraphics{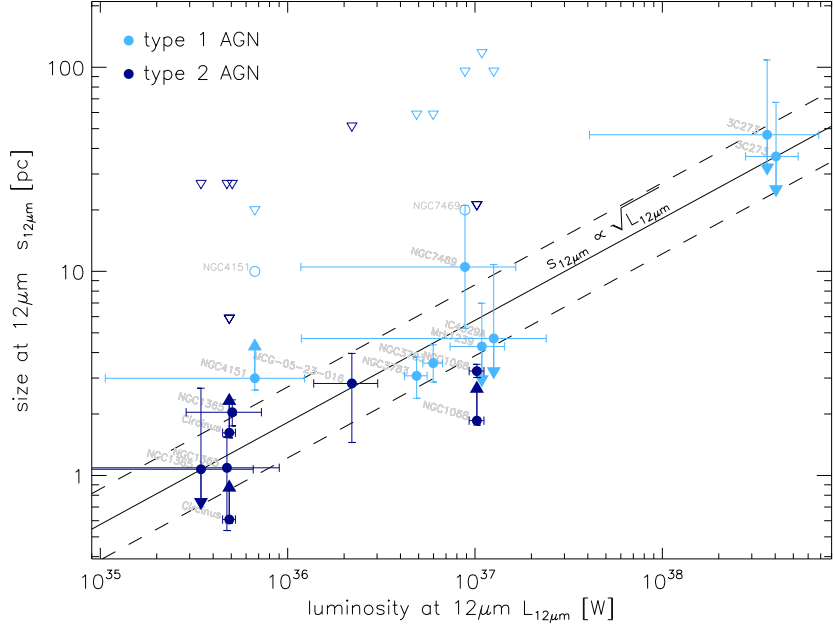}

\caption{Size of the mid-infrared emitter as a function of its monochromatic
luminosity in the mid-infrared for the AGN studied with MIDI (filled
circles with error bars). Upper and lower limits on the size estimates
are marked by arrows. The fitted size--luminosity relation for $p=1.8\cdot10^{-18}\,\mathrm{pc}\cdot\left(\mathrm{W}\right)^{-0.5}$
is traced by the black line. The scatter of the measurements around
this relation is $0.6\,\mathrm{dex}$ (black dashed lines). The physical
scales in the respective galaxies corresponding to an angular resolution
of $0.3\,\mathrm{arcsec}$ (diffraction limit of an $8\,\mathrm{m}$
class telescope at $12\,\mu\mathrm{m}$) are labelled by open triangles.
The size of the resolved emission in NGC~4151 ($10\,\mathrm{pc}$)
from \citet{1990Neugebauer} and in NGC~7469 ($\mathrm{\sim}20\,\mathrm{pc}$)
from \citet{2003Soifer} are shown by open circles. \label{fig:discussion_sizes}}

\end{figure*}

\subsection{Spectral features\label{sub:discussion_spectra}}

The most characteristic feature of AGN in the N band is their absorption
or emission by silicate dust, which leads to a broad absorption trough
or emission bump covering almost the entire wavelength range between
$8$ and $13\,\mu\mathrm{m}$. For typical geometries of AGN tori,
the feature is expected to appear in absorption for type 2 objects
and in emission for type 1 objects. Indeed, for most AGN, including
the sources of the snapshot survey, a silicate absorption or emission
feature is present in high signal to noise Spitzer spectra, even if
the signature is very weak (see e.g. \citealt{2005Hao,2005Weedman,2006Buchanan}).
The MIDI spectra presented here only show indications for silicate
absorption or emission for three sources of the snapshot survey: MCG-05-23-016
(feature in absorption, Seyfert 2 galaxy), Mrk~1239 (feature in emission,
Seyfert 1.5 galaxy) and 3C~273 (also in emission, type 1 quasar).
The feature is relatively weak in all of these three sources, especially
in comparison to the deep absorption features found in the two brighter
Seyfert galaxies studied by MIDI, NGC~1068 and the Circinus galaxy
(both Seyfert 2 galaxies; \citealt{2007Tristram2,2004Jaffe1,2009Raban}).
In fact, it is surprising that the two brightest AGN in the mid-infrared
also show the strongest absorption feature. For the Circinus galaxy
and even more so for NGC~1068, changes in the silicate absorption
depth have been observed between the total flux spectra and the correlated
flux spectra. For the sources of the snapshot survey however, we find
no significant difference in the feature strength between the single
dish measurements (or the spectra used for comparison) and the interferometric
measurements. While we can thus rule out any dramatic changes in the
strength of the feature for the sizes probed by our interferometric
measurements, any small changes may be hidden by the low signal to
noise ratio of the data compared to that of the two bright sources.
It is interesting to note that, for the two type 1 AGN, Mrk~1239
and 3C~273, the peak of the emission feature is located at $11\,\mu\mathrm{m}$.
The peak is thus offset against the maximum at $9.7\,\mu\mathrm{m}$
of the absorption coefficient of interstellar dust, which is dominated
by amorphous Olivine ($[\mathrm{Mg},\mathrm{Fe}]_{2}\mathrm{Si}\mathrm{O}_{4}$,
\citealt{2004Kemper}). The shift of the position of the silicate
feature is best explained by more aluminum-rich silica such as Augite
($\mathrm{Ca}_{2}\mathrm{Al}_{2}\mathrm{Si}\mathrm{O}_{7}$), which
has a higher dissociation temperature and which also seems to be responsible
for the changes in the absorption profile observed for NGC~1068 \citep{2004Jaffe1,2009Raban}.

Finally, we note that there are no signs of the PAH features or the
line emission, which are seen in the spectra used for comparison (especially
for NGC~7469, see Fig.~\ref{fig:result_ngc7469}), in either the
correlated flux spectra or in the total flux spectra. The PAH features
are associated with star formation and are destroyed by the hard UV
radiation in the vicinity of the AGN. Their absence thus indicates
that the regions we are probing are dominated by the radiation field
of the AGN.

\subsection{Sizes of the dust distributions\label{sub:discussion_sizes}}

For simple geometries, the visibility reaches a first minimum when
the fringe spacing $\Lambda=\lambda/\mathit{BL}$ (i.e. the {}``resolution'')
of the interferometer equals the characteristic size of the source.
Therefore, the $\mathit{FWHM}$ of a Gaussian distribution is a good
estimate for the spatial scales over which the bulk of the emission
is distributed, as long as $0.2\lesssim V\lesssim0.8$. For $V\lesssim0.2$,
most of the flux has been resolved out while for $V\gtrsim0.8$ the
bulk of the emission remains unresolved. Only lower or upper limits
on the sizes can be given in these cases. 

One has to be careful not to overinterpret the size values derived
from the interferometric measurements. As shown by models of AGN tori
\citep[e.g.][]{2006Hoenig,2008Schartmann} or revealed by our detailed
studies of the Circinus galaxy and of NGC~1068 \citep{2007Tristram2,2009Raban},
the nuclear mid-infrared emission of AGN is complex, with structures
on a variety of size scales. An interferometric detection with MIDI
merely indicates that the correlated flux that was measured is contained
in a relatively compact region with a size not much more than $\Lambda$.
The size scales sampled by the single dishes used to derive the total
flux spectra are much larger than those probed by the interferometer.
It is not clear from only one (or very few) visibility measurements,
how the flux is actually distributed between the two characteristic
size scales and it is very unlikely that the real brightness distribution
follows a Gaussian dependency. This is especially the case for type
1 AGN, where the infrared flux is possibly also affected by the flux
from the accretion disk, which appears as a point source. The limits
derived from the MIDI data for type 1 AGN are hence not only a measure
for the size of the dust emission alone, but for the combination of
the flux of the torus and of the accretion disk. Because the point
source adds a constant to the visibility, independent of the spatial
frequency (i.e. also the baseline length), further measurements (with
the same position angle but different baseline lengths) must be obtained
in order to disentangle the individual components.

For NGC~1068 and the Circinus galaxy, our simple method to estimate
the size assuming a Gaussian brightness distribution can be verified
in comparison to fitting a model of the emitter to several interferometric
measurements. In the present analysis, the measurements with the shortest
and longest available baselines for these two galaxies were included.
The sizes determined by our simple method (see Table \ref{table:discussion_characteristics})
lie between the sizes of the two components of the models in \citet{2009Raban}
and \citet{2007Tristram2}: $1.5$ to $4.0\,\mathrm{pc}$ for NGC~1068
and $0.4$ to $2.0\,\mathrm{pc}$ for the Circinus galaxy. We thus
conclude that our size estimates are indeed realistic.

In Fig.~\ref{fig:discussion_sizes}, the sizes of the emission regions
are compared to the monochromatic luminosity at $12\,\mu\mathrm{m}$,
which is given by $L_{12\,\mu\mathrm{m}}=4\pi D^{2}\nu F_{\mathrm{tot},12\mathrm{\mu m}}$
with $\nu=2.5\cdot10^{13}\,\mathrm{Hz}$. Studying the sizes as a
function of $L_{12\,\mu\mathrm{m}}$ instead of $F_{\mathrm{tot},12\mathrm{\mu m}}$
breaks any dependency of the size on the distance $D$, except for
selection effects (more distant objects being more luminous). The
physical scales corresponding to an angular resolution of $0.3\,\mathrm{arcsec}$,
which is the diffraction limit of an $8\,\mathrm{m}$ class telescope
at $12\,\mu\mathrm{m}$, are also shown for each galaxy (open triangles).
As the mid-infrared cores of all AGN are essentially unresolved by
single dish observations, these size scales are strict upper limits
for the sizes of the dust distributions.

For similar geometries and volume filling factors, the size of the
torus at a certain wavelength, $s_{\lambda}$, should be a function
of its monochromatic luminosity, $L_{\lambda}$, at this wavelength:
$s_{\lambda}=p_{\lambda}\cdot\left(L_{\lambda}\right)^{0.5}$ or $\log\left(s_{\lambda}/\mathrm{pc}\right)=q_{\lambda}+0.5\cdot\log\left(L_{\lambda}/\mathrm{W}\right)$,
where $p_{\lambda}$ and $q_{\lambda}$ are constants largely depending
on the volume filling factor and/or the optical depth of the torus.
The relation reflects that an object of twice the size has four times
more emitting surface. Indeed most of our measurements agree with
the above relation and for $\lambda=12\,\mu\mathrm{m}$ we find $p_{12\,\mu\mathrm{m}}=\left(1.8\pm0.3\right)\cdot10^{-18}\,\mathrm{pc}\cdot\left(\mathrm{W}\right)^{-0.5}$
and $q_{12\,\mu\mathrm{m}}=-17.7\mp0.2$, respectively. The scatter
of the individual sources around the relation is $0.6\,\mathrm{dex}$
(see Fig.~\ref{fig:discussion_sizes}). It can be largely explained
by the uncertainties in our measurements. However, also differences
in the actual geometry and orientation as well as in the filling factor
of the dust distribution have a significant impact on the apparent
size of the emission region. For an optically thick blackbody emitter
with the size $s$, the luminosity is given as $L_{\lambda}=\pi s^{2}\nu F_{\mathrm{bb}}(\nu,T)$.
This can be rearranged to $s=\left(\pi\nu F_{\mathrm{bb}}(\nu,T)\right)^{-0.5}\cdot\left(L_{\lambda}\right)^{0.5}$.
For $p_{\lambda}=\left(\pi\nu F_{\mathrm{bb}}(\nu,T)\right)^{-0.5}$
this is identical to the above relation. For a temperature of $T=300\,\mathrm{K}$,
taken as an average value for the temperatures suggested by the slopes
of the spectra, we obtain $p_{12\,\mu\mathrm{m}}\sim1.8\cdot10^{-18}\,\mathrm{pc}\cdot\left(\mathrm{W}\right)^{-0.5}$.
This implies that the mid-infrared emission is consistent with originating
in a more or less optically thick and thus compact dust distribution
with an average temperature of $300\,\mathrm{K}$.\textcolor{red}{}

At the first glance, there seems to be a tendency for type 1 objects
to have more extended dust distributions than type 2 objects: while
the nuclear dust distribution in NGC~1068 appears relatively compact
in comparison to all other galaxies, the warm dust in NGC~4151 and
NGC 7469 seems to be more extended than average. Further evidence
in this direction comes from \citet{1990Neugebauer} and \citet{2003Soifer},
who partially resolved the mid-infrared cores in just these two galaxies.
Their size estimates, $10\,\mathrm{pc}$ for NGC~4151 and $(\mathrm{<}13)\times26\,\mathrm{pc}$
for NGC~7469, are also shown in Fig.~\ref{fig:discussion_sizes}
(open circles; averaged to $\mathrm{\sim}20\,\mathrm{pc}$ in the
case of NGC~7469) and suggest even more extended emission regions
than measured by our interferometric measurements. In the case of
NGC~7469, the apparently larger extent may partially be explained
by a contamination with the extended emission from the surrounding
starburst to the single dish spectra, although this contamination
cannot be very large due to the lack of PAH signatures in our total
flux spectrum. In the case of NGC~4151, however, there is no circumnuclear
starburst contribution that could contaminate the single dish measurements.
It seems that the warm dust in this galaxy is more extended for its
luminosity than that of any other galaxy of our sample. This is probably
caused by the emission from dust in the ionisation cones, which is
extended on the scales of $3.5\,\mathrm{arcsec}=230\,\mathrm{pc}$
\citep{2003Radomski}. Thus, it appears that a major fraction of the
mid-infrared emission in this source actually originates in the ionisation
cones and not in the torus itself. Together with the relatively red
spectrum at the centre, it seems that this galaxy is also a special
case according to the standard picture in which most of the mid-infrared
emission is expected to originate in the dusty torus. A more detailed
analysis of the interferometric data, including new observations,
will be necessary to substantiate this finding.

Apart from these indications for individual outliers to the size-luminosity
relation, there is no indication from our data that the size of the
emission region differs significantly between type 1 and type 2 objects.
That means that the sizes of the distributions of warm dust are of
the same order in both types of nuclei. This is in strong support
of the unified picture, where the dust distributions in both AGN types
are assumed to be the same. This finding also agrees well with recent
radiative transfer calculations of clumpy AGN tori, where the appearance
of the tori is relatively similar in the mid-infrared \citep{2008Schartmann,2009Schartmann}.

\subsection{Elongation of the dust emission}

We have tentatively detected an elongated mid-infrared emission region
for three sources: NGC~1365, 3C~273 and NGC~7469. In the latter
case, the elongation has been previously detected by \citet{2003Soifer}.
For NGC~1365, the elongation derived from our measurements is roughly
oriented perpendicularly to the axis of symmetry of the system, which
is defined by the axis of the radio jet or the ionisation cone.

This is, again, in agreement with the expectation in the standard
unified paradigm of AGN and to the results found for the three brighter
AGN (NGC~1068, the Circinus galaxy and Centaurus~A), where dust
disks roughly perpendicular to the axis of symmetry of the systems
were found \citep{2009Raban,2007Tristram2,2007Meisenheimer}. More
measurements are, however, needed to confirm these results.

\section{Conclusions\label{sec:conclusions}}

In the MIDI AGN snapshot survey, high resolution interferometric observations
of ten active galactic nuclei were attempted in the mid-infrared with
MIDI at the VLTI. The goal was to determine, which further AGN, in
addition to the previously investigated sources NGC~1068, the Circinus
galaxy and Centaurus~A, are suitable for studies with MIDI and to
derive first estimates for the sizes of the emission regions in the
mid-infrared.

The observations were performed at the sensitivity limit of MIDI and
of the VLTI. We find that, currently, the main limitation for the
observations is the ability to obtain a stable AO correction and to
obtain a secure total flux (i.e. single dish) measurement. No interferometric
measurements could be carried out for two of the sources targeted,
because the nuclei of these galaxies were too faint and too extended
for the MACAO systems to provide an adaptive optics correction. 

For the remaining eight sources, an interferometric signal could indeed
be detected and for seven of them the signal was strong enough to
derive correlated flux spectra. Together with the three brighter
targets NGC~1068, the Circinus galaxy and Centaurus~A as well as
with two further AGN, NGC~3783 \citep{2008Beckert} and \object{NGC~424}
\citep{2009Hoenig2}, this gives a total number of 12 extragalactic
sources for which fringes could be tracked with MIDI. All of these
objects are worth further investigations and more detailed observations
are underway. 

The silicate feature at $10\,\mu\mathrm{m}$ is detected in the MIDI
data for only 3 of the 10 sources. For these three sources, it provides
evidence for some difference between type 1 and type 2 AGN as expected
in the unified model: the feature appears in emission for one type
1 object and in absorption for two type 2 objects.

For the sources where size estimates for the emission regions could
be determined, we have found that the emission regions have sizes
between $1$ and $10\,\mathrm{pc}$ roughly scaling with $\sqrt{L_{\mathrm{MIR}}}$
and for more or less optically thick emission. This {}``size-luminosity
relation'' is consistent with the emission originating in compact
and warm ($T\sim300\,\mathrm{K}$) dust distributions, which are heated
by the central engines of the AGN. In the spectral features as well
as in their sizes have we found no significant differences between
type 1 and type 2 AGN, indicating that the types of objects have similar
signatures of their dust distributions in the mid-infrared.

Larger differences in size appear between individual objects and we
have found that the individual members of a class differ from each
other more than suggested by the standard picture of the dusty torus.
This is in agreement with recent hydrodynamical models of AGN tori
which show that the mid-infrared properties significantly depend on
the detailed configuration of the non-smooth dust distribution.

\begin{acknowledgements}
We thank the anonymous referee for his many helpful suggestions which
significantly improved the paper. We also thank S. Hönig for providing
us with his VISIR spectrum of MCG-05-13-016 before publication. This
work is based in part on archival data obtained with the Spitzer Space
Telescope, which is operated by the Jet Propulsion Laboratory, California
Institute of Technology under a contract with NASA. The work presented
here includes work for the Ph.D. thesis of KT carried out at the MPIA,
Heidelberg. 
\end{acknowledgements}
\begin{appendix}

\section{Remarks on the observations and the data reduction for the individual
sources\label{sec:observations_individual}}

In the following, the observations and the data reduction for the
individual sources of the snapshot survey are presented in greater
detail. The observation log is given in Table \ref{table:observation-log}.
\begin{sidewaystable*} 

\caption{Observation log for the AGN of the snapshot survey, including the
detector integration times (DIT), the number of frames (NDIT) for
the fringe tracks and photometries as well as the lengths (BL) and
position angles (PA) of the projected baselines. The times, ambient
values and baseline properties relate to the start of the fringe track.}

\label{table:observation-log}\centering

\renewcommand{\footnoterule}{}\begin{tabular}{cccccccccrrll}
\hline 
\hline Date and time & \multicolumn{2}{c}{Object} & DIT & \multicolumn{2}{c}{NDIT} & Airmass & Seeing%
\footnote{From the seeing monitor (DIMM) and at $0.5\,\mathrm{\mu m}$ .%
} & BL & \multicolumn{2}{c}{PA} & \multicolumn{2}{l}{Associated calibrator and comments}\tabularnewline
{[}UTC] &  &  & {[}s] & fringes & phot. &  & {[}$\mathrm{''}$] & {[}m] & \multicolumn{2}{c}{{[}°]} &  & \tabularnewline
\hline
\textbf{2005 Dec 19: } & \textbf{UT2~--~UT3} & \multicolumn{3}{c}{\textbf{076.B-0038(A)}} &  &  &  &  &  &  &  & \tabularnewline
\hline
05:07:40 & HD\,090\,610 & cal01 & 0.018 & 8000 & 2500 & 1.72 & 0.91 & 45.1 & 0.9 &  &  & ok\tabularnewline
06:39:51 & MCG-05-23-016 & sci01 & 0.018 & 12000 & $2\times8000$ & 1.12 & 0.75 & 46.1 & 24.3 &  & cal01 & ok\tabularnewline
07:31:59 & Mrk~1239 & sci02 & 0.018 & 12000 & $2\times8000$ & 1.14 & 0.70 & 41.1 & 36.3 &  & cal02 & ok\tabularnewline
08:15:17 & HD\,083\,618 & cal02 & 0.018 & 8000 & 2500 & 1.09 & 0.51 & 43.9 & 41.7 &  &  & ok\tabularnewline
08:47:23 & Mrk~1239 & sci03 & 0.018 & 12000 & $2\times8000$ & 1.09 & 0.55 & 44.8 & 42.9 &  & cal03 & second part of photometry B contains no signal\tabularnewline
09:27:55 & HD\,083\,618 & cal03 & 0.018 & 8000 &  & 1.13 & 1.00 & 46.3 & 45.3 &  &  & replaced missing photometry by that of cal02\tabularnewline
\textbf{2006 Sep 11:} & \textbf{UT2 -- UT3} & \multicolumn{3}{c}{\textbf{077.B-0026(B)}} &  &  &  &  &  &  &  & \tabularnewline
\hline
05:00:21 & HD\,220\,009 & cal11 & 0.018 & 8000 & 4000 & 1.16 & 1.15 & 44.1 & 45.0 &  &  & ok\tabularnewline
06:09:20 & NGC~7469 & sci11 & 0.018 & 8000 & 8000 & 1.35 & 1.12 & 46.4 & 46.3 &  & cal11 & photometry B contains no signal (AO problem on UT2)\tabularnewline
07:00:40 & HD\,026\,967 & cal12 & 0.018 & 8000 & 4000 & 1.24 & 1.48 & 46.6 & 17.9 &  &  & bumpy fringe track\tabularnewline
07:42:55 & NGC~1365 & sci12 & 0.018 & 8000 & 8000 & 1.06 & 1.31 & 46.6 & 31.1 &  & cal14 & photometry B contains no signal (AO problem on UT2)\tabularnewline
08:28:27 & HD\,026\,967 & cal13 & 0.018 & 8000 &  & 1.08 & 1.32 & 46.4 & 32.1 &  &  & replaced missing photometry by that of cal12\tabularnewline
08:44:20 & HD\,036\,597 & cal14 & 0.018 & 8000 & 4000 & 1.15 & 1.27 & 46.4 & 22.5 &  &  & ok\tabularnewline
09:36:20 & NGC~1365 & sci13 & 0.018 & 8000 &  & 1.03 & 1.37 & 45.1 & 45.7 &  &  & no fringe track, no photometry\tabularnewline
09:41:27 & NGC~1365 & sci14 & 0.018 & 8000 &  & 1.04 &  & 44.9 & 46.3 &  &  & no fringe track (but weak signal), no photometry\tabularnewline
\textbf{2007 Feb 07:} & \textbf{UT2~--~UT3} & \multicolumn{3}{c}{\textbf{078.B-0031(A)}} &  &  &  &  &  &  &  & \tabularnewline
\hline
03:36:01 & HD\,081\,420 & cal21 & 0.018 & 8000 & 4000 & 1.13 & 0.65 & 41.4 & 33.5 &  &  & no offset tracking mode, phot B bad!\tabularnewline
05:41:06 & 3C~273 & sci21 & 0.018 & 8000 &  & 1.37 & 0.69 & 35.2 & 27.0 &  &  & no fringe track (started at wrong OPD position)\tabularnewline
05:50:44 & 3C~273 & sci22 & 0.018 & 8000 &  & 1.33 & 0.83 & 35.8 & 28.7 &  &  & lost fringe track, no photometry\tabularnewline
06:04:10 & 3C~273 & sci23 & 0.018 & 8000 &  & 1.29 & 0.74 & 36.7 & 31.0 &  & cal22 & no photometry\tabularnewline
06:08:32 & 3C~273 & sci24 & 0.018 & 8000 &  & 1.27 & 0.71 & 37.0 & 31.6 &  & cal22 & no photometry\tabularnewline
06:31:50 & HD\,098\,430 & cal22 & 0.018 & 8000 & 4000 & 1.02 & 0.60 & 45.9 & 38.6 &  &  & ok\tabularnewline
07:49:59 & NGC~4151 & sci25 & 0.018 & 8000 &  & 2.27 & 0.67 & 34.6 & 61.9 &  &  & no fringe track (but weak signal), no photometry\tabularnewline
08:03:42 & NGC~4151 & sci26 & 0.018 & 8000 & 4000 (A) & 2.29 & 0.49 & 35.8 & 60.8 &  & cal22 & lost fringe track, only A photometry observed\tabularnewline
08:43:55 & IC~4329A & sci27 & 0.018 & 8000 &  & 1.02 & 0.62 & 46.6 & 35.7 &  & cal23 & no photometry\tabularnewline
08:49:02 & IC~4329A & sci28 & 0.018 & 8000 & 4000 & 1.01 & 0.69 & 46.6 & 36.3 &  & cal23 & photometry observed after cal23\tabularnewline
09:08:18 & HD\,123\,123 & cal23 & 0.018 & 8000 & 4000 & 1.01 & 0.56 & 46.6 & 36.6 &  &  & ok\tabularnewline
\textbf{2007 Nov 24:} & \textbf{UT3~--~UT4} & \multicolumn{3}{c}{\textbf{080.B-0258(A)}} &  &  &  &  &  &  &  & \tabularnewline
\hline
00:00:39 & HD\,220\,009 & cal31 & 0.018 & 8000 & 4000 & 1.16 & 1.15 & 60.9 & 108.1 &  &  & ok\tabularnewline
01:07:41 & NGC~7469 & sci31 & 0.018 & 8000 &  & 1.32 & 1.08 & 51.9 & 106.8 &  & cal31 & no fringe track (but weak signal), IRIS problems on UT3\tabularnewline
01:12:56 & NGC~7469 & sci32 & 0.018 & 8000 &  & 1.34 & 1.17 & 51.1 & 106.9 &  & cal31 & lost fringe track, IRIS problems on UT3, no photometry\tabularnewline
01:53:47 & NGC~1365 & sci33 & 0.018 & 8000 & 8000 & 1.15 & 1.37 & 54.4 & 92.8 &  & cal32 & problems with IRIS\tabularnewline
02:38:26 & HD\,016\,815 & cal32 & 0.018 & 8000 & 2500 & 1.04 & 1.46 & 61.3 & 105.5 &  &  & ok\tabularnewline
03:10:23 & IRAS~05189-2524 & sci34 & 0.018 & 8000 &  & 1.23 &  & 51.1 & 96.0 &  & cal32 & lost track, problems with IRIS, no photometry\tabularnewline
03:33:48 & HD\,016\,815 & cal33 & 0.018 & 8000 &  & 1.04 & 1.11 & 62.5 & 114.0 &  &  & no photometry\tabularnewline
03:51:36 & NGC~1365 & sci35 & 0.018 & 12000 & 8000 & 1.02 & 0.98 & 62.1 & 108.9 &  & cal33 & ok\tabularnewline
04:11:21 & NGC~1365 & sci36 & 0.018 & 12000 &  & 1.02 & 0.98 & 62.4 & 111.8 &  & cal33 & no photometry\tabularnewline
04:18:56 & NGC~1365 & sci37 & 0.018 & 12000 &  & 1.02 & 0.78 & 62.5 & 112.9 &  & cal33 & no photometry\tabularnewline
\textbf{2008 Apr 22:} & \textbf{UT3~--~UT4} & \multicolumn{3}{c}{\textbf{081.D-0092(A)}} &  &  &  &  &  &  &  & \tabularnewline
\hline
06:10:08 & HD\,107\,328 & cal41 & 0.018 & 8000 & 4000 & 1.71 & 0.66 & 36.6 & 118.3 &  &  & ok\tabularnewline
06:52:20 & 3C~273 & sci41 & 0.018 & 8000 & 8000 & 2.04 & 0.69 & 30.7 & 126.6 &  & cal41 & very weak photometry\tabularnewline
07:03:47 & 3C~273 & sci42 & 0.018 & 8000 & 8000 & 2.21 & 0.59 & 28.6 & 129.8 &  & cal41 & photometry B contains no signal\tabularnewline
\hline
\end{tabular}\normalsize

\end{sidewaystable*}

\subsection{NGC~1365\label{sub:observ_ngc1365}}

This source was successfully observed several times during the observation
runs on 2006 Sep 11 and on 2007 Nov 24, using the UT2~--~UT3 and
the UT3~--~UT4 baselines, respectively.

During the first run, the ambient conditions were fair, with a seeing
of $1.1$ to $1.4\,\mathrm{arcsec}$. Although the MACAO unit on UT2
was unable to securely lock on the extended nucleus under these conditions,
a fringe signal could be detected and tracked (sci12, see Table~\ref{table:observation-log}).
While the quality of the subsequent photometry of beam A is relatively
good, no signal is present in the photometry of beam B: no stable
AO correction could be achieved while chopping on UT2. For this reason,
the photometry was replaced by a portion of the A photometry with
little background variations in order to be able to reduce the data
using the standard data reduction routines. Additionally to the standard
reduction routine, which relies on both photometries and yields an
{}``averaged two-dish total spectrum'' of the source, we determined
the total flux using only the photometry of beam A for both NGC~1365
and the associated calibrator HD\,026\,967 (cal12). The information
of beam B was disregarded completely. As for the standard data reduction,
the calibration was carried out using the template spectrum of the
star from the database by Roy van Boekel (private comm.). We find
that the two determinations of the total flux spectrum are consistent
with each other within a fraction of their errors. 

Two further attempts (sci13 and sci14) to track the fringes of this
source on 2007 Nov 24 failed; no stable tracking could be achieved.

More successful observations of the source were carried out during
the second run, although the ambient conditions were comparable to
those on 2006 Sep 11. A total of 4 successful fringe tracks and two
sets of photometric data could be obtained. However, the data reduction
was again not straight forward due to problems with the photometry
of the first data set (sci33): the background in the photometry of
beam A (UT4) was extremely unstable, so that only a subset of the
frames, where the background is relatively stable, were selected for
the data reduction. The last three fringe tracks (sci35, sci36 and
sci37) were observed consecutively without reacquisition of the source
and the length and orientation of the baseline only changed insignificantly.
Therefore, the three measurements were almost identical and they were
averaged to give one visibility point.

\subsection{IRAS~05189-2524\label{sub:observ_leda17155}}

For IRAS~05189-2524 (LEDA~17155), a Seyfert 2 galaxy located at
a distance of $170\,\mathrm{Mpc}$ ($12\,760\pm54\,\mathrm{km}\,\mathrm{s}^{-1}$,
\citealp{1983Huchra}; $1\,\mathrm{arcsec}=850\,\mathrm{pc}$), two
observation attempts were carried out. Although the acquisition succeeded
with UT3 on 2006 Sep 11, it completely failed for UT2. Hence, no data
were obtained on this occasion. Only during the second attempt on
2007 Nov 24 with the UT3~--~UT4 baseline, a stable AO correction
with MACAO was achieved. A very weak signal is present in the fringe
search, although this was only noticed in post processing using the
optimised mask and the parameter settings described in Sect. \ref{sec:observations}.
The fringe track, which was observed {}``blind'', i.e. without knowing
the exact position of zero OPD at the time of the observation, also
shows a very faint signal, while the OPD drifted over the zero OPD
position. However, the signal was too weak for the observing software
to stabilise the OPD and hence not enough frames were recorded to
allow a confident determination of the correlated flux. From the acquisition
images it seems that the two beams were misaligned by more than 2
pixels ($\mathrm{=}0.17\,\mathrm{arcsec}$), which is slightly more
than $\mathit{HWHM}=0.15\,\mathrm{arcsec}$, the half width half maximum
of the PSF averaged over the N band. This means that the beam overlap
was such that only $75\,\%$ of the flux were able to interfere. From
the few frames where an interferometric signal is present, we estimate
that the correlated flux rises from $\mathrm{\sim}0.1\,\mathrm{Jy}$
at $8\,\mathrm{\mu m}$ to $\mathrm{\sim}0.2\,\mathrm{Jy}$ at $13\,\mathrm{\mu m}$.
Taking into account the correction for the beam overlap, the correlated
flux in IRAS~05189-2524 seems to have been on the order of $0.1$
to $0.3\,\mathrm{Jy}$. This is significantly less than the total
flux spectrum measured by \citet{2004Siebenmorgen}, which rises from
$0.3$ to $1.0\,\mathrm{Jy}$ from the short to the long wavelength
end of the N band, potentially indicating that the source is resolved.
Unfortunately, no photometry confirming this conclusion was observed
for IRAS~05189-2524 with MIDI. This case shows how important it is
to also obtain photometric information even for those sources where
no fringes could be detected or tracked. Then a lower limit for the
size of the emitter could be derived.

\subsection{MCG-05-23-016 and Mrk~1239 \label{sub:observ_mrk-and-mcg}}

One full visibility measurement for MCG-05-23-016 and two full measurements
for Mrk~1239 were obtained on 2005 Dec 19. Double photometric measurements
of 8000 frames each for every visibility point were carried out in
order to increase the signal-to-noise and to obtain redundant measurements
in case of problems. Indeed, for the second observation of Mrk~1239
(sci03, see Table~\ref{table:observation-log}), the MACAO loop was
open for a portion of the second photometry of beam B (UT2). The faulty
photometry was replaced by the corresponding one taken during the
first observation (sci02, see Table \ref{table:observation-log}).

The second observation of the calibrator HD\,083\,618 was obtained
at the end of the night and no further photometric data could be recorded.
The data set was completed using the photometry from the first observation
of this calibrator (cal02).

To obtain the final fluxes and the visibility, the two individual
measurements of Mrk~1239 were averaged. This is possible because
the position angle and baseline length only differ insignificantly.

\subsection{NGC~3281\label{sub:observ_ngc3281}}

An unsuccessful attempt to observe the Seyfert 2 galaxy NGC~3281
($v_{\mathrm{hel}}=3200\pm22\,\mathrm{km}\,\mathrm{s}^{-1}$, \citealp{1998Theureau};
$D=45\,\mathrm{Mpc}$, $1\,\mathrm{arcsec}=210\,\mathrm{pc}$) was
made during the observing run on 2007 Feb 07. MACAO could not lock
on the nucleus of this galaxy because it was too extended in the visible.
As a consequence, the source could also not be detected with IRIS
and no data were obtained at all.

\subsection{NGC~4151\label{sub:observ_ngc4151}}

Our first observation of NGC~4151 was carried out on 2007 Feb 07
using the UT2~--~UT3 baseline. From the two fringe tracks that were
attempted, only the second one contains useful data: the fringes were
clearly tracked for about $40\,\mathrm{sec}$ before they were lost.
Only these $40\,\mathrm{sec}$ of tracking were used for the determination
of the correlated flux. Due to the failure of MACAO to keep the loop
closed on UT2 while chopping, only the photometry for beam A contains
useful data. A pipeline reduction with EWS was performed using the
A photometry for beam B. The result was double checked by an additional
determination of the total flux only using the A photometries of NGC~4151
and of HD\,098\,430 alone, as described for the first observation
of NGC~1365 (see Appendix~\ref{sub:observ_ngc1365}): the two total
flux spectra agree within $30\,\%$.

More successful observations were carried out on 2008 Apr 22 and 2008
Apr 24 with the UT3~--~UT4 and the UT2~--~UT4 baselines, respectively.
These new observations as well as a full analysis for the interferometric
data will be presented in a separate paper, \citealt{2009Burtscher}.

\subsection{3C~273\label{sub:observ_3c273}}

3C~273 was first observed during the observation run on 2007 Feb
07. MACAO could not keep the loop closed during chopping and, as a
result, no acquisition images and photometry were obtained. Given
the pointlike nucleus with $V=12.8\,\mathrm{mag}$ \citep{2006Turler}
and the later success closing the loop on the nucleus (see below),
the reason for this malfunction remains unexplained. Instead, the
beams were aligned with IRIS in the K band alone. In total, four attempts
to track fringes were performed, however only the last two contain
an interferometric signal. During the data reduction they were treated
as a single, long track. Because no photometric data were available,
only the calibrated correlated flux could be determined from the MIDI
data. To obtain a rough estimate for the total flux of the source,
intermediate band photometry with VISIR was obtained on 2007 Mar 01,
that is, 23 days after the interferometric observations. VISIR, the
VLT Imager and Spectrometer for the mid-InfraRed, provides a long-slit
spectrometer as well as a high sensitivity imager in both the N and
the Q bands \citep{2004Lagage}. The two filters, in which 3C~273
was observed, were PAH1 ($\lambda_{0}=8.59\,\mu\mathrm{m}$, $\Delta\lambda=0.42\,\mu\mathrm{m}$)
and SiC ($\lambda_{0}=11.85\,\mu\mathrm{m}$, $\Delta\lambda=2.34\,\mu\mathrm{m}$;
\citealp{2007ESOMAN-VISIR}). The calibrator star for the VISIR observations
was HD\,124\,294. The VISIR data were reduced by combining the chopped
and nodded frames in the standard procedures for such data. The photometry
was extracted in a $2.25\,\mathrm{arcsec}$ aperture. For the calibration,
the template spectrum of HD\,124\,294 from the catalogue by Roy
van Boekel (private comm\emph{.}) was used. The errors were estimated
by increasing the aperture to $3.00\,\mathrm{arcsec}$ and by decreasing
it to $1.50\,\mathrm{arcsec}$ -- for both the calibrator and 3C~273
in the opposite way. In the VISIR images most of the flux is contained
within $1.5\,\mathrm{arcsec}$ and the FWHM of the PSF is $0.39\,\mathrm{arcsec}$
and $0.44\,\mathrm{arcsec}$ at $8.59\,\mu\mathrm{m}$ and $11.85\,\mu\mathrm{m}$,
respectively. This is only slightly larger than the FWHM of the calibrator
star. Therefore, 3C~273 is not resolved by the VISIR imaging.

More successful observations of 3C~273 were obtained on 2008 Apr
22 with the UT3~--~UT4 baseline. Two data sets (fringe track plus
photometry) were observed directly one after the other in order to
increase the signal to noise ratio. They were both reduced together
because no significant change in baseline length and orientation took
place between the measurements. The photometric measurements have
a very low signal to noise ratio. This is mainly due to very strong
variations in the background emission that are insufficiently removed
by the chopping. In order to get non-negative fluxes for photometry
A from the first data set (sci41, see Table~\ref{table:observation-log}),
a special removal of the background was carried out by fitting a 5th
order polynomial to the sky emission on both sides of the spectrum.
During the reduction of the photometry, we noticed that different
total flux spectra can be obtained by changing the mask and the position
of the sky bands. The alternative results are well represented by
the error bars of the photometry from EWS.

\subsection{IC~4329A\label{sub:observ_ic4329a}}

This target was also observed during the run on 2007 Feb 07. At first,
the MACAO units on both telescopes could not keep the loop closed
while chopping. During the interferometric measurement, however, a
stable AO correction was possible. Strong fringes were found and tracked
in two separate fringe tracks (sci27 and sci28, see Table~\ref{table:observation-log}).
These were combined and reduced as one data set. Only later after
the observation of the corresponding calibrator, HD\,123\,123 (cal23),
a stable AO correction could be achieved also while chopping. The
photometry was thus observed in twilight, 1 hour after the fringe
tracks. Because no proper alignment of the beams was carried out for
the photometry, these are located at different positions on the detector
than those of the interferometric signal. By consequence, the masks
had to be adjusted accordingly. Still, the fluxes measured in the
two beams differ by more than a factor of 2 and must hence be considered
with caution.

\subsection{NGC~5506\label{sub:observ_5506}}

For the galaxy NGC~5506, which contains a Seyfert 1.9 nucleus and
is located at a distance of $25\,\mathrm{Mpc}$ ($v_{\mathrm{hel}}=1850\pm8\,\mathrm{km}\,\mathrm{s}^{-1}$,
\citealp{1996Keel}; $1\,\mathrm{arcsec}=120\,\mathrm{pc}$), observations
were attempted on 2007 Feb 07. However, on both telescopes (UT2 and
UT3) MACAO could not lock on the extended and faint nucleus. Hence
no data were obtained.

\subsection{NGC~7469\label{sub:observ_ngc7469}}

Two attempts to observe this source were carried out, the first one
on 2006 Sep 11 with the baseline UT2~--~UT3, the second one on 2007
Nov 24 with the baseline UT3~--~UT4.

During the first attempt, a fringe track (sci11, see Table~\ref{table:observation-log})
was successfully performed, although the signal was extremely weak
and the position of the OPD was not always determined correctly. As
for NGC~1365, there was no AO correction by MACAO for the subsequent
observation of the photometry of beam B (UT2). As a result, the single
dish spectrum is very extended spatially and it is unclear to what
degree it is dominated by background fluctuations. For instance, no
pronounced ozone absorption feature at $9.7\,\mu\mathrm{m}$ is present
in the raw data, indicating that most of the signal is actually spurious.
We therefore only selected those portions of the photometry for the
data reduction where the background residual after chopping is relatively
flat. Additionally, the total flux spectrum of the source was only
determined by using the information from beam A of the source and
that of the associated calibrator HD\,220\,009 (cal11), as described
in Appendix~\ref{sub:observ_ngc1365}.

During the second attempt to observe NGC~7469, the environmental
conditions were similar to those of the first attempt, with moderate
wind and the seeing varying between $1.1\,\mathrm{arcsec}$ and $1.3\,\mathrm{arcsec}$.
This time, however, MACAO obtained a stable AO correction on the nucleus.
The source was successfully centered in MIDI after switching from
the SiC to the N8.7 filter and increasing the chopping frequency to
$2.3\,\mathrm{Hz}$%
\footnote{The chopping frequency was set to a non-integer number in order to
not have the same frequency as the gas-helium closed-cycle cooler
of MIDI which runs at $1\,\mathrm{Hz}$. This should cancel out any
differences in the background between target and sky position, which
may be caused by a slight temperature change or vibrations with the
frequency of the cooler.%
}. Very faint fringes could be detected in the fringe search, however
these were too faint to assure a stable fringe tracking: in both of
the two tracks that were performed, the OPD drifted away from the
zero OPD value. From the few frames where an interferometric signal
is present, it seems that the correlated flux was on the order of
$0.1$ to $0.2\,\mathrm{Jy}$, rising towards longer wavelengths.
No photometry was observed on this occasion.

\end{appendix}

\bibliographystyle{aa}
\bibliography{11607mas}

\end{document}